\pdfoutput=1
\documentclass[
  aps,
  prl,
  reprint,
  superscriptaddress,
  nofootinbib,
	longbibliography
]{revtex4-2}

\usepackage{amsmath,amssymb}
\usepackage{bm}
\usepackage{graphicx}
\usepackage[percent]{overpic}
\usepackage{dcolumn}
\usepackage[version=4]{mhchem}
\usepackage[percent]{overpic}
\usepackage{siunitx}
\usepackage{xr}
\usepackage[labeled,resetlabels]{multibib}
\usepackage{etoolbox}
\newcites{S}{Supplemental References}

\makeatletter
\AtBeginDocument{%
  \def\NAT@bibsetnum#1{%
    \setlength{\topsep}{\z@}%
    \NATx@bibsetnum{#1}%
  }%
  \let\@bibsetup\NAT@bibsetnum
  \patchcmd{\endthebibliography}%
    {\label{LastBibItem}}%
    {}%
    {}%
    {\PackageWarning{bibfix}{Could not patch bibliography}}%
}
\makeatother

\usepackage{color}
\usepackage[colorlinks,bookmarks=false,citecolor=darkblue,linkcolor=red,urlcolor=blue]{hyperref}

\definecolor{darkred}{rgb}{0.7,0.0,0.0}
\definecolor{darkblue}{rgb}{0,0.02,0.45}

\begin{document}

\title{Beyond Kitaev exchange: $\Gamma'$-driven magnetism in Ag$_3$LiRh$_2$O$_6$}

\author{Efrain Insuasti Pazmino}
\altaffiliation{These authors contributed equally to this work}
\affiliation{Felix Bloch Institute for Solid-State Physics, University of
Leipzig, 04103 Leipzig, Germany}

\author{Shi Feng}
\altaffiliation{These authors contributed equally to this work}
\affiliation{Technical University of Munich, TUM School of Natural Sciences, Physics Department, 85748 Garching, Germany}
\affiliation{Munich Center for Quantum Science and Technology (MCQST), Schellingstr. 4, 80799 M{\"u}nchen, Germany}

\author{Ramesh Dhakal}
\affiliation{Department of Physics and Center for Functional Materials, Wake
Forest University, Winston-Salem, North Carolina 27109, USA}

\author{Ioannis Rousochatzakis}
\affiliation{Department of Physics, Loughborough University, Loughborough LE11 3TU, United Kingdom}

\author{Stephen M. Winter}
\email{winters@wfu.edu}
\affiliation{Department of Physics and Center for Functional Materials, Wake
Forest University, Winston-Salem, North Carolina 27109, USA}

\author{Johannes Knolle}
\email{j.knolle@tum.de}
\affiliation{Technical University of Munich, TUM School of Natural Sciences, Physics Department, 85748 Garching, Germany}
\affiliation{Munich Center for Quantum Science and Technology (MCQST), Schellingstr. 4, 80799 M{\"u}nchen, Germany}
\affiliation{Blackett Laboratory, Imperial College London, London SW7 2AZ, United Kingdom}

\author{Alexander A. Tsirlin}
\email{altsirlin@gmail.com}
\affiliation{Felix Bloch Institute for Solid-State Physics, University of
Leipzig, 04103 Leipzig, Germany}

\date{\today}

\begin{abstract}
In Kitaev magnets, frustration arises from bond-dependent spin anisotropic interactions, which can potentially realize sought-after quantum spin liquid phases. However, material realizations with dominating Kitaev exchange are scarce and the broad parameter space of extended Kitaev models remains largely unexplored experimentally. Here, we use density-functional theory, cluster many-body, spin-wave, and density-matrix renormalization calculations to show that the recently discovered honeycomb rhodate Ag$_3$LiRh$_2$O$_6$ hosts an off-diagonal exchange $\Gamma_1'$ that is second in magnitude only to the ferromagnetic Kitaev exchange $K_1$. This combination of $K_1$ and $\Gamma_1'$ explains key experimental observations, an unusually high magnetic ordering temperature of 94\,K at ambient pressure and its rapid decrease under hydrostatic pressure as Ag$_3$LiRh$_2$O$_6$ evolves from ferromagnetic toward zigzag order within the honeycomb planes. Our spin-wave calculations further predict an unusual topology of the magnon bands arising from the $\Gamma_1'$ term in the spin Hamiltonian of the honeycomb layer. We show that the large trigonal splitting of $t_{2g}$ states enhances $\Gamma_1'$, the hitherto underappreciated term in Kitaev magnets, and opens the way to new interesting regimes of the extended Kitaev model.
\end{abstract}

\maketitle

\begin{figure}[t]
  \centering
  \hspace*{1cm}
  \begin{overpic}[width=0.85\columnwidth]{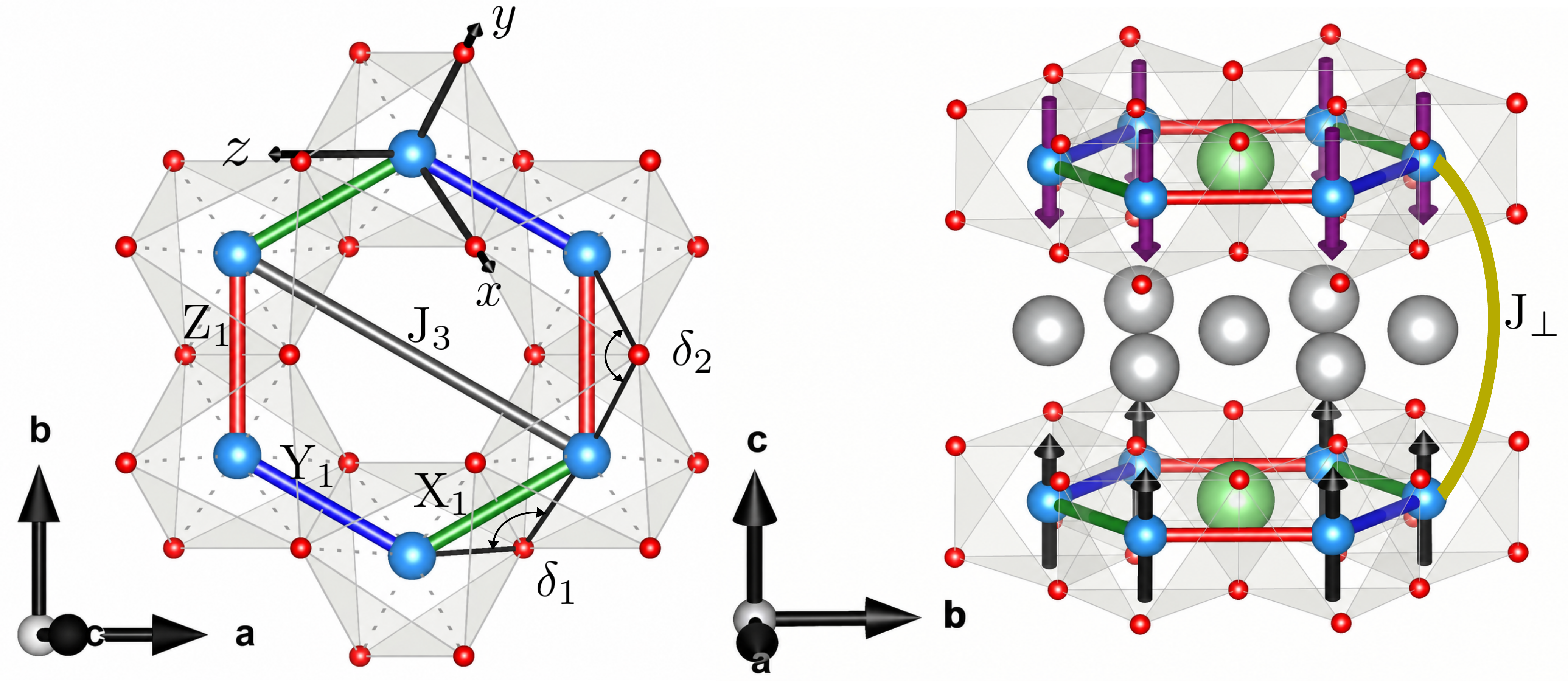}
    \put(-14,38){\small (\textit{a})}
  \end{overpic}
  \includegraphics[width=\columnwidth]{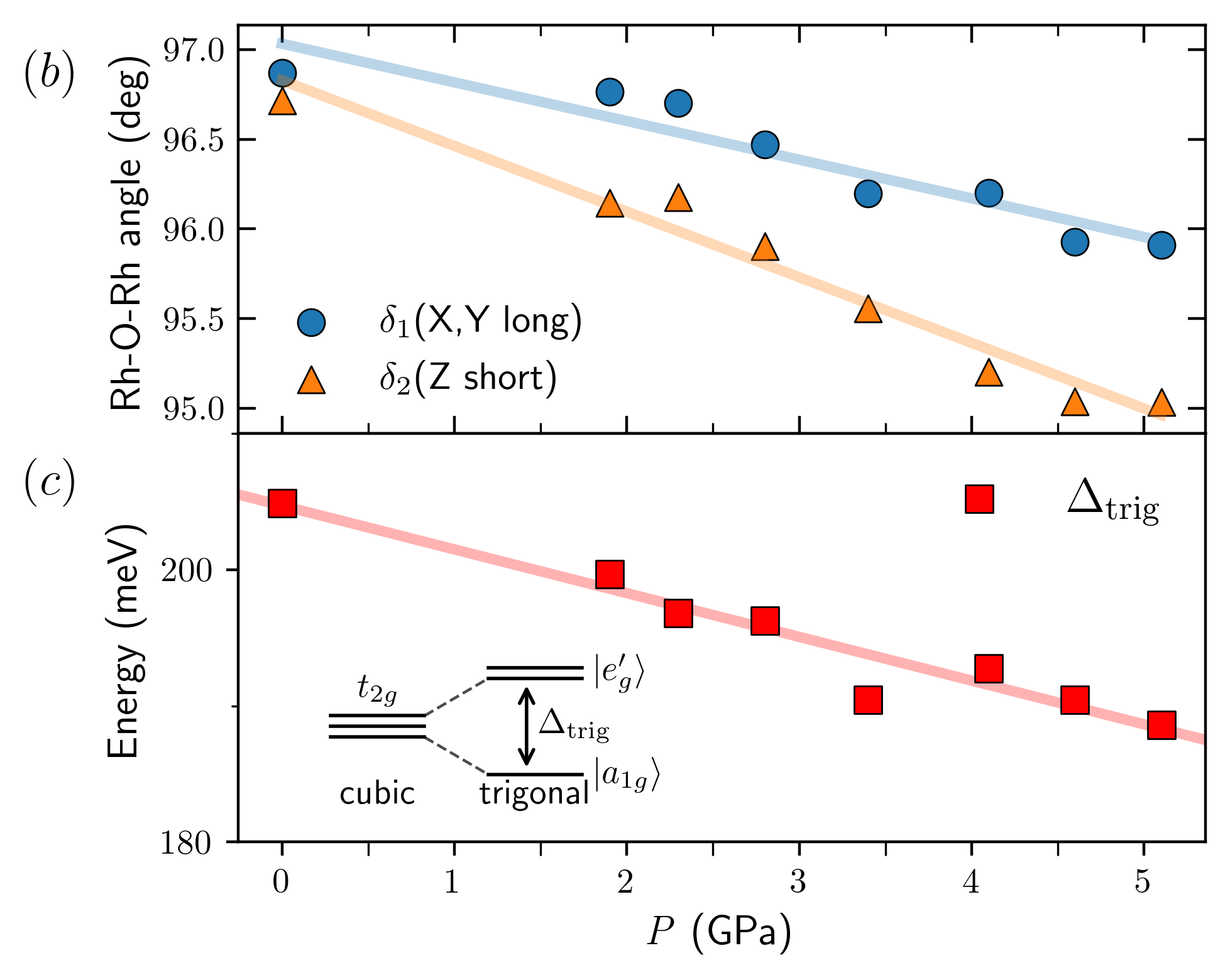}
  \caption{\label{fig:rh-o-rh-angle}
  (a)~Left: honeycomb plane with bond-anisotropic exchange couplings and the relevant bond angles $\delta_1$ and $\delta_2$. Right: Proposed A-type AFM order along with the interlayer coupling $J_\perp$. 
  (b)~Pressure dependence of the Rh--O--Rh bond angles for the long $X,Y$
  bonds ($\delta_1$) and the short $Z$ bonds ($\delta_2$).
  (c)~Pressure dependence of the trigonal crystal-field splitting shown in the inset.
  All solid lines are guides to the eye. }
\end{figure}

\textit{Introduction.---}
The Kitaev model of interacting spins on the honeycomb lattice opened a new direction in magnetism research by introducing an original mechanism of frustration driven by exchange anisotropy~\cite{kitaev2006}. This model provides a rare example of an analytically solvable quantum spin liquid ground state whose excitations are described by two types of emergent degrees of freedom, fluxes and Majorana fermions~\cite{hermanns2018,takagi2019,trebst2022}. Bond-anisotropic Kitaev exchange, the central ingredient of this model, occurs in real materials, but always in combination with other exchange terms~\cite{winter2017}. In honeycomb magnets, the nearest-neighbor exchange tensor usually takes the form 
\begin{equation}
    \mathbb{J}_Z =
    \begin{pmatrix}
      J_1 & \Gamma_1 & \Gamma_1' \\
      \Gamma_1 & J_1 & \Gamma_1' \\
      \Gamma_1' & \Gamma_1' & J_1+K_1
    \end{pmatrix},
  \label{eq:exchange-tensor}
\end{equation}
for the $Z$-bonds of the hexagon (Fig.~\ref{fig:rh-o-rh-angle}a), whereas the $X$- and $Y$- bonds are obtained by permutations. Here, $J_1$ is the Heisenberg coupling, and $K_1$ is the Kitaev exchange of the eponymous model, the subscript $1$ denotes nearest neighbors. Two other terms, the off-diagonal anisotropies $\Gamma_1$ and $\Gamma_1'$, are of importance too because they originate from the same hopping processes as the Kitaev exchange and always accompany it in real materials~\cite{winter2016}. The typical Jackeli-Khaliullin scenario of $j_{\rm eff}=\frac12$ magnetic ions located within edge-shared MO$_6$ octahedra (Fig.~\ref{fig:rh-o-rh-angle}a) relies on the large ligand-mediated hopping that gives rise to ferromagnetic (FM) $K_1$ along with a sizable positive $\Gamma_1$, whereas $J_1$ and especially $\Gamma_1'$ are usually smaller. This regime has been now established in a variety of Kitaev candidates, including $\alpha$-RuCl$_3$~\cite{matsuda2025,moeller2025} and BaCo$_2$(AsO$_4$)$_2$~\cite{maksimov2025}. 

The resulting minimal model, $K_1-\Gamma_1$ on the honeycomb lattice, exhibits strong frustration spanning the range between a quantum spin liquid at large $K_1$ and classical spin liquid at large $\Gamma_1$~\cite{rousochatzakis2017,rousochatzakis2024}. Experimentally, frustration is usually alleviated by weak third-neighbor antiferromagnetic exchange that stabilizes zigzag or related types of order, but such an order appears to be rather fragile. It sets in below $T_N$ of $5-15$\,K to be compared with the typical energy scale of $50-100$\,K for the leading exchange term $K_1$~\cite{winter2017}. A rapid suppression of this order under in-plane magnetic fields opens the possibility of stabilizing a disordered ground state at finite fields, although the relation of such a field-induced state to the Kitaev spin liquid remains vividly debated~\cite{matsuda2025,feng2026}.

An alternative approach toward Kitaev spin liquid requires material exploration in a broader range of parameters of the extended $J_1-K_1-\Gamma_1-\Gamma_1'$ Kitaev model~\cite{rousochatzakis2024} beyond the $K_1-\Gamma_1$ line where the majority of existing Kitaev candidates fall in. One interesting material in this context is the honeycomb rhodate Ag$_3$LiRh$_2$O$_6$, see Fig.~\ref{fig:rh-o-rh-angle}(a), which is magnetically ordered with $T_N=94$\,K~\cite{bahrami2022}, showing a remarkably high transition temperature among the Kitaev candidates. This $T_N$ is rapidly suppressed under hydrostatic pressure, with increased spin fluctuations witnessed by muon spectroscopy~\cite{sakrikar2025}. High-pressure magnetometry does not detect any magnetic ordering transition above 4\,GPa~\cite{sakrikar2025}, suggesting the possibility of a pressure-tuned magnetically disordered ground state, as opposed to the field-tuned quantum-disordered state in other Kitaev candidates.

Despite this interesting and very distinct phenomenology, the microscopic scenario of Ag$_3$LiRh$_2$O$_6$ remains unexplored, and even the ambient-pressure magnetic ground state of this material was never reported. Here, we fill this gap by combining \textit{ab initio} and cluster many-body calculations with tensor-network methods to determine the microscopic regime of Ag$_3$LiRh$_2$O$_6$ and elucidate its tunability under pressure. We show that Ag$_3$LiRh$_2$O$_6$ lies in the previously unexplored parameter range of the extended Kitaev model where ferromagnetic $K_1$ is accompanied by negative off-diagonal $\Gamma_1'$ as the main subleading term. The $\Gamma_1$ term is smaller in magnitude than $\Gamma_1'$, yet pressure tuning of the $\Gamma_1/K_1$ ratio at constant $\Gamma_1'/K_1$ allows Ag$_3$LiRh$_2$O$_6$ to traverse the phase boundary between FM and zigzag orders. As the first real-world material with the dominant Kitaev term, sizable $\Gamma_1'$, and FM in-plane order, Ag$_3$LiRh$_2$O$_6$ is also interesting in the context of the unusual magnon band topology.

\textit{Methods.---}
Magnetic couplings in Ag$_3$LiRh$_2$O$_6$ were obtained by cluster many-body calculations for a two-site Hubbard model~\cite{riedl2019} with the hopping parameters determined by density-functional theory (DFT) calculations for crystal structures at various pressures. Experimental lattice parameters were used as input, whereas atomic positions were relaxed. The quantum phase diagram was obtained from the ground states of density-matrix renormalization group (DMRG) calculations on a YC4 cylinder (8-site circumference) with fixed bond dimension $1600$ and truncation error $\sim 10^{-9}$. We then employed linear spin-wave theory (LSWT) to analyze the spectrum and topology of magnon bands for the 2D spin model. Further details can be found in the Supplemental Material~\cite{SM}.

\textit{Ambient-pressure behavior.---}
Ag$_3$LiRh$_2$O$_6$ is a derivative of the simple honeycomb rhodate Li$_2$RhO$_3$ prepared via topotactic ion exchange of interlayer Li ions by Ag~\cite{bahrami2022}. This synthesis method leaves the structure mostly unchanged, although the unit-cell volume increases by 35\% from 217.8\,\r A$^3$ in Li$_2$RhO$_3$~\cite{todorova2011} to 295.4\,\r A$^3$ in Ag$_3$LiRh$_2$O$_6$~\cite{bahrami2022}, creating an effective negative pressure. The monoclinic crystal structure ($C2/m$) entails layers of edge-shared RhO$_6$ octahedra with honeycombs centered by Li atoms. This structural geometry results in Rh--O--Rh bond angles ($\varphi$) close to $90^{\circ}$. In Ag$_3$LiRh$_2$O$_6$ ($\varphi\simeq 96.8^{\circ}$), the deviation from $90^{\circ}$ appears to be larger than in Li$_2$RhO$_3$ and other Kitaev candidates where bond angles usually lie in the range of $92-94^{\circ}$. Another distinctive feature of the material is its sizable trigonal distortion of the RhO$_6$ octahedra, mirrored by the unusually low x-ray absorption branching ratio of 2.22, to be compared with 3.17 in Li$_2$RhO$_3$ and 4.0 as the atomic value for the $j_{\rm eff}=\frac12$ state~\cite{bahrami2022}. This is in contrast to materials like $\alpha$-RuCl$_3$ where an almost pure $j_{\rm eff}=\frac12$ state of the magnetic ion has been established~\cite{agrestini2017}.

Existing experimental evidence strongly suggests that replacing interlayer Li with Ag not only creates negative chemical pressure but also modifies the electronic state of the Rh$^{4+}$ ion. We now explore the implications of this effect for the magnetic couplings. Our cluster many-body calculations identify the nearest-neighbor FM Kitaev exchange, $\bar K_1\simeq -8.4$\,meV, as the largest coupling in magnitude. Here, the bar denotes an average over the $X$-, $Y$-, and $Z$-bonds, which are inequivalent in the monoclinic crystal structure but feature similar couplings~\cite{SM}. Surprisingly, the second-largest coupling is $\bar\Gamma_1'\simeq -4.7$\,meV, rather than $\bar\Gamma_1\simeq -1.8$\,meV. The magnitude of $\bar\Gamma_1$ is instead comparable to that of the antiferromagnetic third-neighbor coupling, $\bar J_3\simeq 1.9$\,meV. Moreover, $\bar\Gamma_1$ is negative, in contrast to the sizable positive $\Gamma_1$ characteristic of the $K_1-\Gamma_1$ regime of $\alpha$-RuCl$_3$ and other Kitaev magnets.

\begin{figure}[t]
  \centering
  \includegraphics[width=\columnwidth]{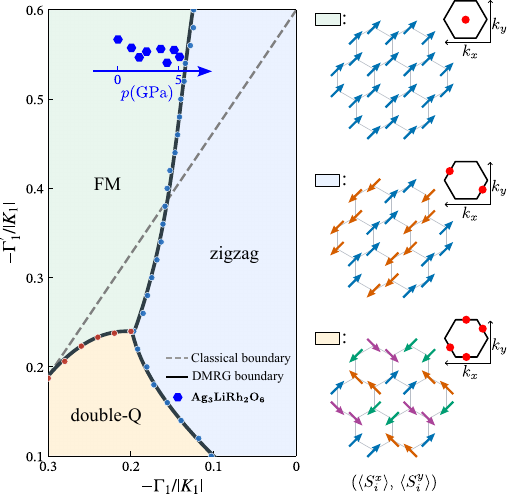}
  \caption{\label{fig:phase}
  Phase diagram of the monolayer $K_1-\Gamma_1-\Gamma_1^\prime-J_3$ model as a function of the normalized interactions $-\Gamma_1/|K_1|$ and $-\Gamma_1'/|K_1|$, with $J_3/|K_1| = 0.2$. Solid and dashed curves mark the DMRG and classical phase boundaries; red and blue dots on the boundaries denote the phase transitions derived from susceptibilities with respect to $\Gamma_1^\prime$ and $\Gamma_1$; Blue hexagonal markers in the FM phase show the CED parameters under compression. Real-space spin ordering along with static structure factors on the right column are shown for the ferromagnetic (FM), zigzag, and the double-$Q$ phases respectively. }
\end{figure}

The minimum microscopic magnetic model of Ag$_3$LiRh$_2$O$_6$ is thus $K_1-\Gamma_1'$ with the addition of $\Gamma_1<0$ and $J_3$ as weaker terms. Its simplified version, the $K_1-\Gamma_1'$ model, is known to stabilize FM in-plane order for $K_1<0$ and $\Gamma_1'<0$~\cite{yilmaz2022}. Antiferromagnetic $J_3$ frustrates such an order, but our DMRG simulations confirm that FM order is retained even in the presence of $J_3$ for the relevant set of parameters (Fig.~\ref{fig:phase}). At first glance, the proposed FM order contradicts the apparent antiferromagnetism of Ag$_3$LiRh$_2$O$_6$~\cite{bahrami2022}. This discrepancy is resolved by taking into account the interlayer coupling $J_{\perp}$, which is antiferromagnetic. Our calculations thus predict A-type antiferromagnetic order in Ag$_3$LiRh$_2$O$_6$, with ferromagnetic honeycomb layers stacked antiferromagnetically, as illustrated in the right panel of Fig.~\ref{fig:rh-o-rh-angle}(a). Simulation of the neutron diffraction pattern~\cite{SM} shows that such an order produces magnetic Bragg peaks only outside of the $Q$-range probed experimentally in Ref.~\cite{sakrikar2025} where no magnetic scattering could be observed.

\begin{figure}[t]
  \centering
  \includegraphics[width=\columnwidth]{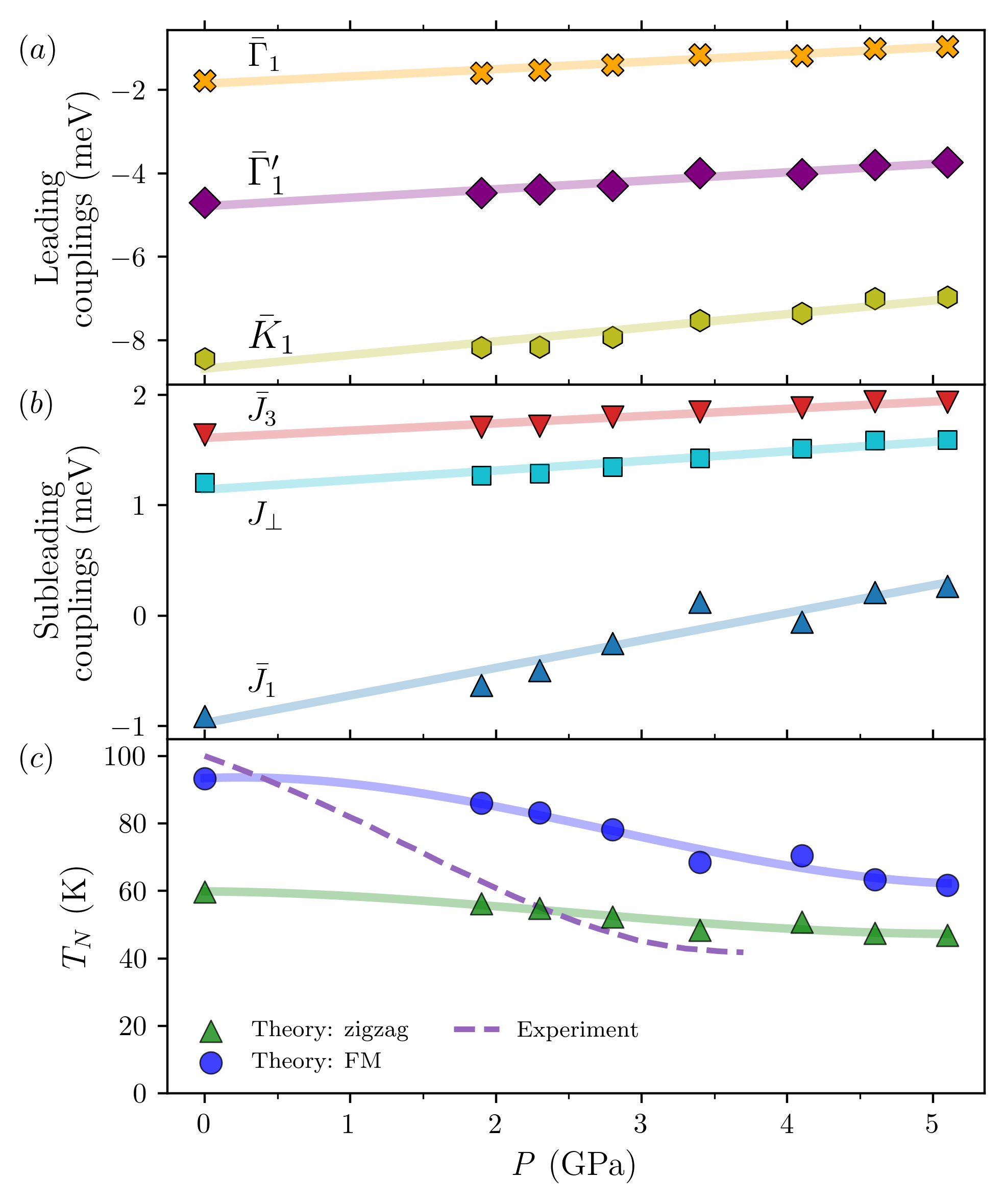}
  \caption{\label{fig:exchange-couplings}
  Pressure evolution of the exchange couplings obtained from cluster exact
  diagonalization (CED) and mean-field estimate of the Neél temperature: (a) Leading couplings $\bar{K}_1$, $\bar{\Gamma}'_1$ and $\bar{\Gamma}_1$. (b) $\bar{J}_3$ and $J_{\perp}$ constitute the subleading terms, while $\bar{J}_1$ is rather small and gets even weaker under pressure.
  (c) Mean-field calculation of $T_N$. At lower pressures, the A-type AFM order reproduces the experimental values, whereas at higher pressures a better agreement is obtained for the zigzag order.} 
\end{figure}

The scenario of in-plane ferromagnetism elucidates the magnetic behavior of Ag$_3$LiRh$_2$O$_6$. This material has an unusually high $T_N=94$\,K, whereas above $T_N$ its magnetic susceptibility almost perfectly follows the Curie-Weiss law showing no evidence of short-range order~\cite{bahrami2022}. Moreover, our mean-field estimate of $T_N$~\cite{SM} returns values of about 90\,K in good agreement with the experiment, see Fig.~\ref{fig:exchange-couplings}(c). All of this indicates that quantum effects are weak in this material at ambient pressure, because negative $\Gamma_1'$ strongly favors ferromagnetic order and quenches quantum fluctuations. 
This is consistent with our spin-wave analysis of the in-plane FM state, which yields a large magnon gap of the order $\Delta_0/|K_1|=\mathcal{O}(1)$, as discussed in the next sections.

Qualitative differences between Ag$_3$LiRh$_2$O$_6$ and other Kitaev candidates (zigzag vs. FM in-plane order, low vs. high $T_N$) are thus a consequence of the sizable $\Gamma_1'$ term. This enhanced $\Gamma_1'$ can be partially attributed to the increased deviation of the Rh--O--Rh bond angle from $90^{\circ}$ \cite{UnpubRau2014}. Indeed, a comparable value of $\bar\Gamma_1'\simeq -3.0$\,meV was obtained in \textit{ab initio} calculations for Na$_2$IrO$_3$ with $\varphi=98-100^{\circ}$, but in that case the effect of $\Gamma_1'$ is overshadowed by $K_1$ and $J_3$, which are both larger in magnitude~\cite{winter2016}. 

Trigonal distortion of the RhO$_6$ octahedra is another important ingredient. Our DFT calculations yield the trigonal splitting of about 200\,meV that exceeds the typical spin-orbit coupling of $\lambda=150$\,meV in rhodates, see Fig.~\ref{fig:rh-o-rh-angle}(c), and corroborates major deviations from the $j_{\rm eff}=\frac12$ state, in accordance with the x-ray absorption branching ratios measured experimentally~\cite{bahrami2022}. The effect of the trigonal distortion on the magnetism of Ag$_3$LiRh$_2$O$_6$ can be quantified by extracting exchange couplings from the second-order perturbation theory~\cite{winter2016} that uses DFT hopping parameters and takes into account the full structural geometry, but neglects the effect of the trigonal distortion on the electronic state of Rh$^{4+}$. Such calculation returns $\bar\Gamma_1'\simeq -1.6$\,meV, thrice lower in magnitude compared to the exact value, and a weakly positive $\bar\Gamma_1\simeq 0.7$\,meV. These parameters would put Ag$_3$LiRh$_2$O$_6$ deep into the region of the zigzag order on the phase diagram.

\textit{Pressure evolution.---} 
Most of the Kitaev candidates are rather insensitive to hydrostatic pressure and demonstrate only a weak change in $T_N$ before reaching structural phase transitions accompanied by magnetism collapse~\cite{bastien2018,biesner2018,hermann2018,clancy2018,veiga2019,shen2021,shen2024,jin2024}. A similar structural phase transition has been observed in Ag$_3$LiRh$_2$O$_6$ around 6.6\,GPa, but below this pressure the material shows a remarkable tunability with $dT_N/dP\simeq -20$\,K/GPa~\cite{sakrikar2025} to be compared with +1.4\,K/GPa in $\alpha$-Li$_2$IrO$_3$~\cite{shen2022}. This behavior can now be elucidated using our microscopic magnetic model. 

Our calculations show that the Rh--O--Rh bond angle, as well as the trigonal distortion of the RhO$_6$ octahedra, gradually decrease under pressure [Fig.~\ref{fig:rh-o-rh-angle}(b)]. This weakens $\bar K_1$, $\bar\Gamma_1'$, and $\bar\Gamma_1$ [Fig.~\ref{fig:exchange-couplings}(a)]; whereas $J_3$ and $J_{\perp}$ slightly increase with pressure [Fig.~\ref{fig:exchange-couplings}(b)]. The $\bar\Gamma_1'/\bar K_1$ ratio remains essentially unchanged, but the $\bar\Gamma_1/\bar K_1$ ratio systematically decreases and serves as a microscopic tuning parameter. 

To uncover the influence of this tuning parameter on the ground state, we used DMRG to construct a quantum phase diagram of the $K_1-\Gamma_1-\Gamma_1'-J_3$ model with $K_1<0$ and $\Gamma_1'<0$, shown in Fig.~\ref{fig:phase}, as relevant to Ag$_3$LiRh$_2$O$_6$. We find ferromagnetic order stabilized at large $\Gamma_1'/K_1$ and sufficiently negative $\Gamma_1$, whereas weakly negative $\Gamma_1$ allows $J_3$ to stabilize zigzag order. The reason for this competition is that $\Gamma_1<0$ and $\Gamma_1'<0$ both lower the energy of the FM state when the spins point along the [111] direction in the Kitaev $XYZ$-frame, i.e., perpendicular to the honeycomb layers. On the other hand, $\Gamma_1>0$, as typical for most of the Kitaev candidates, frustrates FM order.

Pressure tuning of Ag$_3$LiRh$_2$O$_6$ mainly changes $\Gamma_1/K_1$ and shifts the material from the region of the in-plane FM order toward the boundary with the zigzag state, which may be reached around 5\,GPa, close to the threshold of the structural phase transition. Interestingly, quantum effects in Ag$_3$LiRh$_2$O$_6$ should be enhanced under pressure. Classical calculations reveal a qualitatively similar phase boundary between the FM and zigzag states, but the exact position of this boundary is significantly modified when quantum fluctuations are taken into account by DMRG, see Fig.~\ref{fig:phase}. 
Note that although the FM–zigzag transition is found to be near $\Gamma_1/|K_1|\simeq-0.15$ from DMRG and near $\Gamma_1/|K_1|\simeq-0.05$ from classical approximation, the linear spin-wave expansion about the FM reference retains a finite excitation gap across both predictions. The persistence of a metastable FM branch is consistent with a first-order transition between the FM and the zigzag states. 
Another observation at this juncture is that the $T_N$ of Ag$_3$LiRh$_2$O$_6$ increasingly deviates from the mean-field prediction as pressure increases. All of this is consistent with the progressive evolution of the material toward the FM-zigzag phase boundary. Linear extrapolation of the experimental $dT_N/dP\simeq -20$\,K/GPa would suggest that $T_N$ drops to zero around 5\,GPa in a very good agreement with our observation that at 5\,GPa Ag$_3$LiRh$_2$O$_6$ lies almost on the phase boundary. This microscopic scenario explains the remarkable tunability of the material, though its reduction in $T_N$ under pressure should be interpreted as an evolution toward another ordered state instead of spin liquid.
Finally, we point out that at smaller $|\Gamma_1'|$, the DMRG phase diagram contains a noncollinear double-$Q$ phase (yellow in Fig.~\ref{fig:phase}). It combines modulations at two symmetry-related wave vectors into a multi-$Q$ spin texture~\cite{Yuchen24}, see also Supplemental Material \cite{SM}. This type of order intriguingly resembles the double-$Q$ order recently found in Na$_3$Co$_2$SbO$_6$~\cite{Yuchen24}, but further analysis of this phase lies beyond the scope of the present work.

\begin{figure}[b]
  \centering
  \includegraphics[width=\columnwidth]{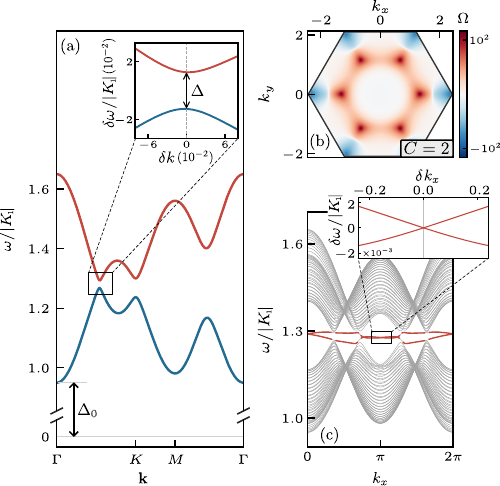}
  \caption{\label{fig:lsw} Zero-field magnon spectrum for $S=\frac{1}{2}$ at $(K_1,\Gamma_1,\Gamma_1',J_3)=(-1,-0.3,-0.4,0.2)$ within the FM phase. (a) Bulk magnon bands. Inset: zoom-in of the avoided crossing at $k_x = \pi$. (b) Lower-band Berry curvature $\Omega$ in log scale. (c) Band structure on a cylinder with two zigzag-edge modes' zero-energy crossing (in red) at $k_x = 0,~\pi$; the inset shows the zoomed-in edge mode at $k_x = \pi$.
  }
\end{figure}

\textit{Magnetic excitations.---} 
We now turn to a theoretical prediction of magnon bands in the FM layers of Ag$_3$LiRh$_2$O$_6$. Rotating Eq.~\eqref{eq:exchange-tensor} into the local spin frame and expanding in Holstein--Primakoff bosons along the easy axis gives the magnon bilinear $H_{\rm LSW}=\sum_{\bm k}\Psi_{\bm k}^\dagger\mathcal H(\bm k)\Psi_{\bm k}$, where $\Psi_{\bm k}=(a_{A\bm k},a_{B\bm k},a^\dagger_{A,-\bm k},a^\dagger_{B,-\bm k})^{\mathsf T}$ resolves the two honeycomb sublattices \cite{Karaki23,Inhee26}. Technical details are given in the Supplemental Material~\cite{SM}. 

Figure~\ref{fig:lsw}(a) shows the two positive-energy magnon bands at a representative point inside the FM phase. The spectrum has no gapless Goldstone mode due to the absence of a continuous spin-rotation symmetry, and instead exhibits a large magnon gap $\Delta_0/|K_1| \sim \mathcal{O}(1)$ at the $\rm \Gamma$ point. The two bands are also separated by a smaller gap $\Delta$ forming an avoided crossing. The sizable magnon gap $\Delta_0$ indicates low-energy transverse fluctuations of the FM moment are strongly suppressed at ambient pressure; indeed, we find that $\langle\delta S\rangle/S$ is less than $5\%$ in the FM phase even for $S = \frac{1}{2}$, which is consistent with the experimental observation of the almost classical behavior at zero pressure.

Remarkably, the magnon bands with a finite avoided crossing $\Delta$ [Fig.~\ref{fig:lsw}(a)] in the FM phase are topologically nontrivial. Each magnon band carries a finite Berry curvature, concentrated predominantly near the avoided crossings between $\rm K$ and $\rm \Gamma$ points [Fig.~\ref{fig:lsw}(b)], and its integral gives the Chern number $|C|=2$. 
The corresponding bulk--edge correspondence is visible in the cylinder spectrum of Fig.~\ref{fig:lsw}(c). Two edge-localized branches traverse the bulk gap and cross at $k_x=0$ and $\pi$. 
The large Berry curvature implies an intrinsic magnon thermal Hall conductivity $\kappa_{xy}$ for a single FM layer \cite{Matsumoto14}. This response is thermally activated across $\Delta_0$, and is expected to become appreciable when magnons near the Berry-curvature hot spots are populated, and reverses sign upon reversal of the layer magnetization. Although in the bulk A-type antiferromagnet, contributions from oppositely ordered layers may cancel at zero field, a field-polarized or layer-imbalanced state should display a finite thermal Hall conductivity, which is open for future study.

\textit{Discussion.---}
Among Kitaev candidates, Ag$_3$LiRh$_2$O$_6$ is an outlier because of its high and strongly pressure-dependent $T_N$. Interestingly, pressure evolution of this material follows general trends established across various Ir$^{4+}$, Ru$^{3+}$, and Rh$^{4+}$ compounds~\cite{winter2016,tsirlin2022,shen2024,shen2025}. Pressure reduces the bond angle $\varphi$ and weakens the FM Kitaev exchange $K_1$, while shifting $\Gamma_1$ in the positive direction. In fact, even the chemical pressure effect upon the Li by Ag substitution can be broadly described within the same framework. Negative pressure increases $\varphi$ by about $2^{\circ}$ in Ag$_3$LiRh$_2$O$_6$ compared to Li$_2$RhO$_3$ and switches $\Gamma_1$ from positive to negative. The distinctive feature of Ag$_3$LiRh$_2$O$_6$ is its large trigonal distortion that competes with the spin-orbit coupling and results in a sizable negative $\Gamma_1'$, the hitherto underappreciated term in the study of extended Kitaev models. This term is particularly important at low pressures. Compression leads to a gradual symmetrization of the structure, reduction in $|\Gamma_1'|$, and the evolution toward zigzag order, the common ground state of several Kitaev magnets.

The $\Gamma_1'$ term has two-fold influence on the magnetism. On the one hand, it reduces frustration and stabilizes in-plane FM order with the high ordering temperature and little quantum effects, although quantum fluctuations may be enhanced upon approaching the FM-zigzag phase boundary. On the other hand, the FM state itself is by no means mundane, as it features $|C|=2$ topological magnons. Unlike the known $|C|=1$ magnon phases in $J_1-K_1-\Gamma_1$ magnets \cite{Joshi18,McClarty18} or the nodal magnon phase in the polarized Kitaev model \cite{Nasu18,Liang18,Feng24}, the boundary of each FM plane hosts two co-propagating chiral modes, producing a multichannel edge spectrum and a potentially enhanced thermal Hall response due to large magnon Berry curvature. Alternatively the topological magnons can be probed by energy-resolved spectroscopy with Raman circular dichroism~\cite{vinas2023direct,koller2025raman}.

The enhancement of $\Gamma_1'$ and the realization of the hitherto unexplored parameter regime of the extended Kitaev model is a joint effect of the bond angle $\varphi$ deviating from $90^{\circ}$ and the electronic state of Rh$^{4+}$ deviating from $j_{\rm eff}=\frac12$, owing to the large trigonal distortion. Similar bond angles of $\varphi\simeq 96^{\circ}$ have been reported in the iridate analog, Ag$_3$LiIr$_2$O$_6$~\cite{bahrami2019,bahrami2021}, but the microscopic regime of that compound~\cite{li2022} largely follows the scenario of its parent Li$_2$IrO$_3$ (positive $\Gamma_1$, small $\Gamma_1'$), likely because in a $5d$ magnet the trigonal splitting rarely exceeds the spin-orbit coupling. The weaker spin-orbit coupling in $4d$ magnets amplifies changes caused by chemical substitution. This renders rhodates a particularly fruitful playground for reaching new regimes of the extended Kitaev model, including FM states with topological magnons.

\textit{Conclusion.---}
Ag$_3$LiRh$_2$O$_6$ establishes a new regime of Kitaev magnetism in which the off-diagonal anisotropy $\Gamma_1'$ is not a secondary correction, but a key ingredient shaping both the magnetic ground state and its pressure evolution. The negative $\Gamma_1'$ stabilizes ferromagnetic in-plane order and gives rise to gapped spin-wave excitations with nontrivial magnon-band topology. Hydrostatic pressure selectively tunes the weaker off-diagonal anisotropy $\Gamma_1$, thereby driving a rapid suppression of $T_N$, enhancing quantum fluctuations, and steering the system toward zigzag order. These results identify Ag$_3$LiRh$_2$O$_6$ as a platform for accessing previously unexplored parameter regimes of the extended Kitaev model and demonstrate that the interplay of trigonal crystal field and spin-orbit coupling in $4d$ honeycomb magnets provides a powerful route to engineering unconventional exchange hierarchies and emergent magnetic states.

\begin{acknowledgments}
\textit{Acknowledgments.---} 
We thank Fazel Tafti for bringing Ag$_3$LiRh$_2$O$_6$ to our attention, and Bin Shen and Philipp Gegenwart for pursuing pressure studies on this material. This work was funded by the Deutsche Forschungsgemeinschaft (DFG, German Research Foundation) -- TRR 360 -- 492547816 (subprojects B1 and B6).
S.F. and J.K. further acknowledge support under Germany’s Excellence Strategy--EXC--2111--390814868, DFG grants No. KN1254/1-2, KN1254/2-1, the Imperial-TUM flagship partnership, the European Research Council (ERC) under the European Union’s Horizon 2020 research and innovation programme (grant agreement No. 851161), the European Union (grant agreement No. 101169765), as well as the Munich Quantum Valley, which is supported by the Bavarian state government with funds from the Hightech Agenda Bayern Plus. Research at Wake Forest University is supported by the National Science Foundation under Grant Number DMR-2338704. DFT calculations were done using resources of the Leipzig University Computing Center, and the Center for Information Services and High Performance Computing (ZIH), Dresden.
\end{acknowledgments}


%

\clearpage

\renewcommand{\citenumfont}[1]{S#1}
\renewcommand{\thesection}{S\Roman{section}}

\setcounter{figure}{0}
\setcounter{table}{0}
\setcounter{equation}{0}

\renewcommand{\thefigure}{S\arabic{figure}}
\renewcommand{\thetable}{S\arabic{table}}
\renewcommand{\theequation}{S\arabic{equation}}

\renewcommand{\theHfigure}{supplement.figure.\arabic{figure}}
\renewcommand{\theHtable}{supplement.table.\arabic{table}}
\renewcommand{\theHequation}{supplement.equation.\arabic{equation}}


\begin{widetext}
\begin{center}
	\textbf{\large \textit{Supplemental Material}\medskip \\
	Beyond Kitaev exchange: $\Gamma'$-driven magnetism in Ag$_3$LiRh$_2$O$_6$}
\end{center}

\section{Density-functional calculations}
\label{sec:sm-dft}
Geometry optimizations were performed using the Vienna Ab initio Simulation Package (VASP) \citeS{SMvasp1,SMvasp2} with PAW-PBE pseudopotentials~\citeS{SMBloechl1994,SMpaw2} and Perdew--Burke--Ernzerhof (PBE) flavor of the exchange-correlation potential~\citeS{SMpbe96}. A plane-wave kinetic energy cutoff of 500 eV was used. The Brillouin zone was sampled with an $8 \times 6 \times 8$ $\Gamma$-centered Monkhorst-Pack k-point grid. Spin-orbit coupling (SOC) and non-collinear magnetism were included self-consistently. To account for strong on-site Coulomb correlations, the DFT+$U$ approach of Dudarev et al. \citeS{SMDudarev1998} was applied to the Rh $d$-orbitals with an effective Hubbard parameter $U_{\rm eff}=U-J=2.29$\, eV. Electronic self-consistent cycles were converged to $10^{-6}$ eV, and Gaussian smearing ($\sigma=0.01$ eV) was used for the electronic occupations. Structural relaxations employed the conjugate-gradient algorithm with fixed cell shape and volume, according to the experimental values at different pressures. Ionic positions were optimized until the residual Hellmann--Feynman forces on all atoms fell below 0.01 eV/\AA.

Band structures and hopping parameters within the Rh $t_{2g}$ shell were obtained in FPLO~\citeS{SMfplo} with the PBE exchange-correlation potential and a $12\times12\times12$ k-point mesh. Wannier functions were calculated using the built-in routine of FPLO, and the resulting tight-binding model was used to calculate exchange couplings on the level of exact second-order perturbation theory (ESOPT). For the cluster exact diagonalization (CED) method, the Wannier functions were obtained based on full-relativistic band structures, using non-spin-polarized PBE+SO calculations.  

\section{Determination of the Trigonal field}
Previous works pointed out the existence of a trigonal distortion $\Delta_{\mathrm{trig}}$ along the [111] direction in the cubic frame~\citeS{SMbahrami2022,SMSakrikar2025}. Therefore, we calculated the trigonal field of the Rh octahedra using orbital energies from DFT+$U$ for the following wave functions \citeS{SMYang2010,SMBhattacharjee_2012}:
\begin{equation}
\begin{aligned}
|a_{1g}\rangle &= \frac{1}{\sqrt{3}}\left(n_1|d_{yz}\rangle + n_2|d_{zx}\rangle + n_3|d_{xy}\rangle\right),\\
|e'_{g1}\rangle &= \frac{1}{\sqrt{3}}\left(\omega n_1|d_{yz}\rangle + \omega^2 n_2|d_{zx}\rangle + n_3|d_{xy}\rangle\right),\\
|e'_{g2}\rangle &= \frac{1}{\sqrt{3}}\left(\omega^2 n_1|d_{yz}\rangle + \omega n_2|d_{zx}\rangle + n_3|d_{xy}\rangle\right).
\end{aligned}
\end{equation}
where $\omega=e^{i2\pi/3}$ and $\mathbf{n}=[1,1,1]$. We found the degenerate levels $|e'_{g1}\rangle$ and $|e'_{g2}\rangle$, which are higher in energy than the lower-lying singlet $|a_{1g}\rangle$, already signaling a trigonal compression \citeS{SMKhomskii_2014}. The trigonal field is then obtained from the expectation values $\Delta_\text{trig}= \langle e'_{g1}|H_\text{orb}|e'_{g1}\rangle - \langle a_{1g}|H_\text{orb}|a_{1g}\rangle$. Furthermore, this $\Delta_\text{trig}$ definition can be compared with the parameters $\Delta_1$ and $\Delta_2$ given in Ref.~\citeS{SMwinter2016} using
\begin{equation}
	-\Delta_\text{trig}=\Delta_1+2\Delta_2.
\end{equation}

A trigonal compression ($\Delta_\text{trig}>0$) is found at all pressures. At 0 GPa, $\Delta_\text{trig} \approx 205$ meV, and its value gets reduced to $\approx 189$ meV at 5.1 GPa (see Fig.~\ref{fig:rh-o-rh-angle}(c) of the main text). Using plain DFT calculations, one finds a smaller $\Delta_\text{trig} \approx 119$ meV at 0 GPa, but the inclusion of the Hubbard $U$ term is expected to produce more accurate results for a correlated material.

\section{Cluster many-body calculations and exchange couplings}
\label{sec:sm-exchange}
In order to evaluate magnetic couplings, we first use the Exact Second-Order Perturbation Theory (ESOPT), constructed over the tight-binding Hamiltonian from Wannier functions \citeS{SMwinter2016} for the magnetic ion \ce{Rh^{4+}}. This methodology considers the ideal case of a perfect octahedral environment without trigonal fields or virtual hoppings between the $t_{2g}$ and $e_g$ manifolds. The resulting couplings are expected to be less accurate than in the case of Cluster Exact Diagonalization (CED), but they provide a useful reference point and highlight the effect of the trigonal crystal-field splitting in Ag$_3$LiRh$_2$O$_6$. 

The goal of both approaches is to obtain a spin Hamiltonian
\begin{equation}
        \mathcal{H}_{\text{spin}} =\sum_{\langle i,j \rangle} \mathbf{S}_i \, \mathbb{J}_{ij} \, \mathbf{S}_j
\label{eq:couptens}
\end{equation}
where $\mathbf{S}_{i}$ and $\mathbf{S}_{j}$ are pseudospin-$\frac{1}{2}$ operators, and $\mathbb{J}_{ij}$ is a $3\times3$ exchange tensor for an arbitrary symmetry, parametrized via:

\begin{equation}
\mathbb{J}_{ij}=\begin{pmatrix}
J_{ij}+\Gamma_{aa} & \Gamma_{ab}+D_{ij}^c & \Gamma_{ac}-D_{ij}^b \\
\Gamma_{ab}-D_{ij}^c &J_{ij}+ \Gamma_{bb} & \Gamma_{bc}+D_{ij}^a \\
\Gamma_{ac}+D_{ij}^b & \Gamma_{bc}-D_{ij}^a & J_{ij}+\Gamma_{cc}
\end{pmatrix}
\label{eqDMV}
\end{equation}
It can be written as:
\begin{equation}
\mathcal{H}_{\text{spin}} = \sum_{\langle ij \rangle} J_{ij} \cdot \mathbf{S}_i \cdot \mathbf{S}_j + \mathbf{S}_i \cdot  \mathbf{\Gamma}_{ij} \cdot \mathbf{S}_j+\mathbf{D}_{ij} \cdot \mathbf{S}_i \times \mathbf{S}_j. \label{eq:equation_20}
\end{equation}
where $J_{ij}$ corresponds to the scalar Heisenberg coupling, $\mathbf{\Gamma}_{ij}$ is a traceless tensor representing the pseudodipolar interaction, $\mathbf{D}_{ij}$ is the Dzyaloshinskii-Moriya vector (which vanishes whenever an inversion center between the atoms exists).
This expression can be related to the spin Hamiltonian of the Kitaev model \citeS{SMkitaev2006} $\mathcal{H} = \sum_{\langle ij\rangle}S_i^\gamma S_j^\gamma$ by choosing a suitable parametrization, which leads to the $J-K-\Gamma-\Gamma'$ model\citeS{SMrau2014,SMwinter2016,SMYamaji2014} (as defined in Fig.~~\ref{fig:rh-o-rh-angle}(a)):
\begin{equation}
\mathcal{H} = \sum_{\langle ij \rangle} \left [J_{ij} \mathbf{S}_i \cdot \mathbf{S}_j + K_{ij} S_i^\gamma S_j^\gamma + \Gamma_{ij}(S_i^\alpha S_j^\beta + S_i^\beta S^\alpha_j)
+ \Gamma_{ij}' ({S}_i^\gamma {S}_j^\alpha + {S}_i^\gamma {S}_j^\beta +{S}_i^\alpha {S}_j ^\gamma + {S}_i^\beta {S}_j^\gamma)\right ]
\end{equation}
Here, $K_{ij}$, corresponds to the Kitaev coupling, and $\Gamma_{ij}$ and $\Gamma_{ij}^\prime$ are the off-diagonal anisotropies, and the super indices $(\alpha, \beta, \gamma)=(x,y,z)$ for the Z-bond can be permuted to obtain the X or Y bonds. All of these interactions are bond-dependent: the $\text{Z}$-bond being the most symmetric one, whereas the $\text{X,Y}$-bonds require additional $\xi$ and $\zeta$ terms for their complete description.
The projection coordinates for the $d$-orbital Wannier functions were chosen such that the cubic axes fulfill the condition $(x+y+z) \perp$ to the $(a, b)$-plane.   For the on-site Coulomb interaction and Hund's coupling we take $U=\SI{2.00}{\electronvolt}$, $J_H=\SI{0.23}{\electronvolt}$. For the ESOPT method $\lambda =\SI{0.15}{\electronvolt}$ was used \citeS{SMeichstaedt2019deriving}. 

The exact diagonalization of the electronic Hamiltonian was carried out on two-site clusters to obtain the low-energy eigenstates. By following the approach of des Cloizeaux Effective Hamiltonians (dCEHs)~\citeS{SMdes1960extension}, these eigenstates are projected onto idealized spin-orbital states to obtain the intermediate basis. This intermediate basis is then orthonormalized following the symmetric (Löwdin) approach \citeS{SMlowdin1950non}.
The complete, non-averaged set of calculated couplings using both methods can be seen in Figs.~\ref{fig:sm-1nn-esopt-ced} and~\ref{fig:sm-3nn-esopt-ced} for the first and third nearest neighbors, respectively.

\begin{figure}[p]
  \centering
  \includegraphics[width=\textwidth]{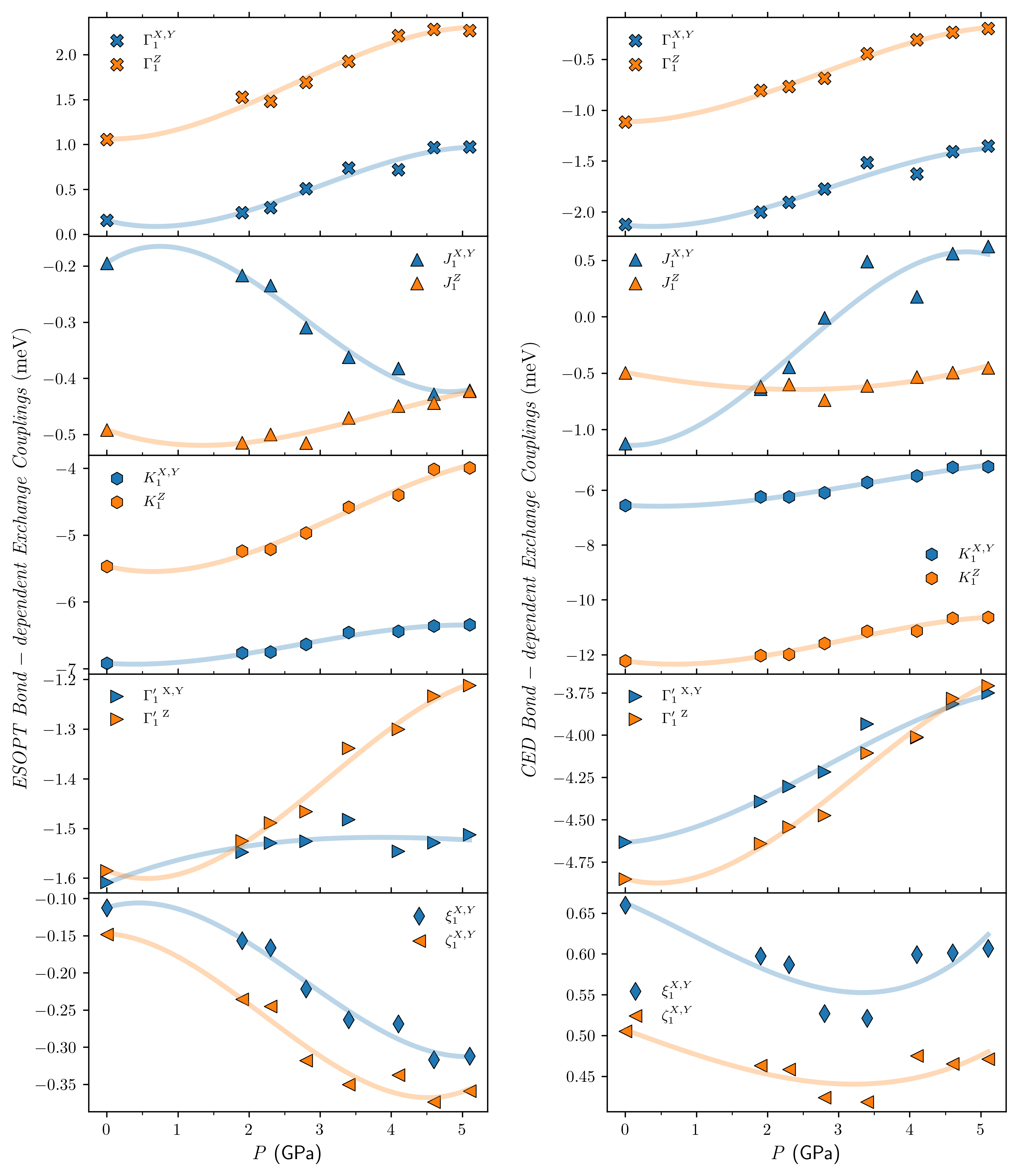}
  \caption{\label{fig:sm-1nn-esopt-ced}
  Pressure dependence of the complete set of nearest-neighbor exchange
  couplings obtained using exact second-order perturbation theory (ESOPT,
  left) and cluster exact diagonalization (CED, right). Blue and orange
  symbols denote the $X,Y$- and $Z$-bond values, respectively; solid curves
  are guides to the eye.}
\end{figure}

\begin{figure}[p]
  \centering
  \includegraphics[width=\textwidth]{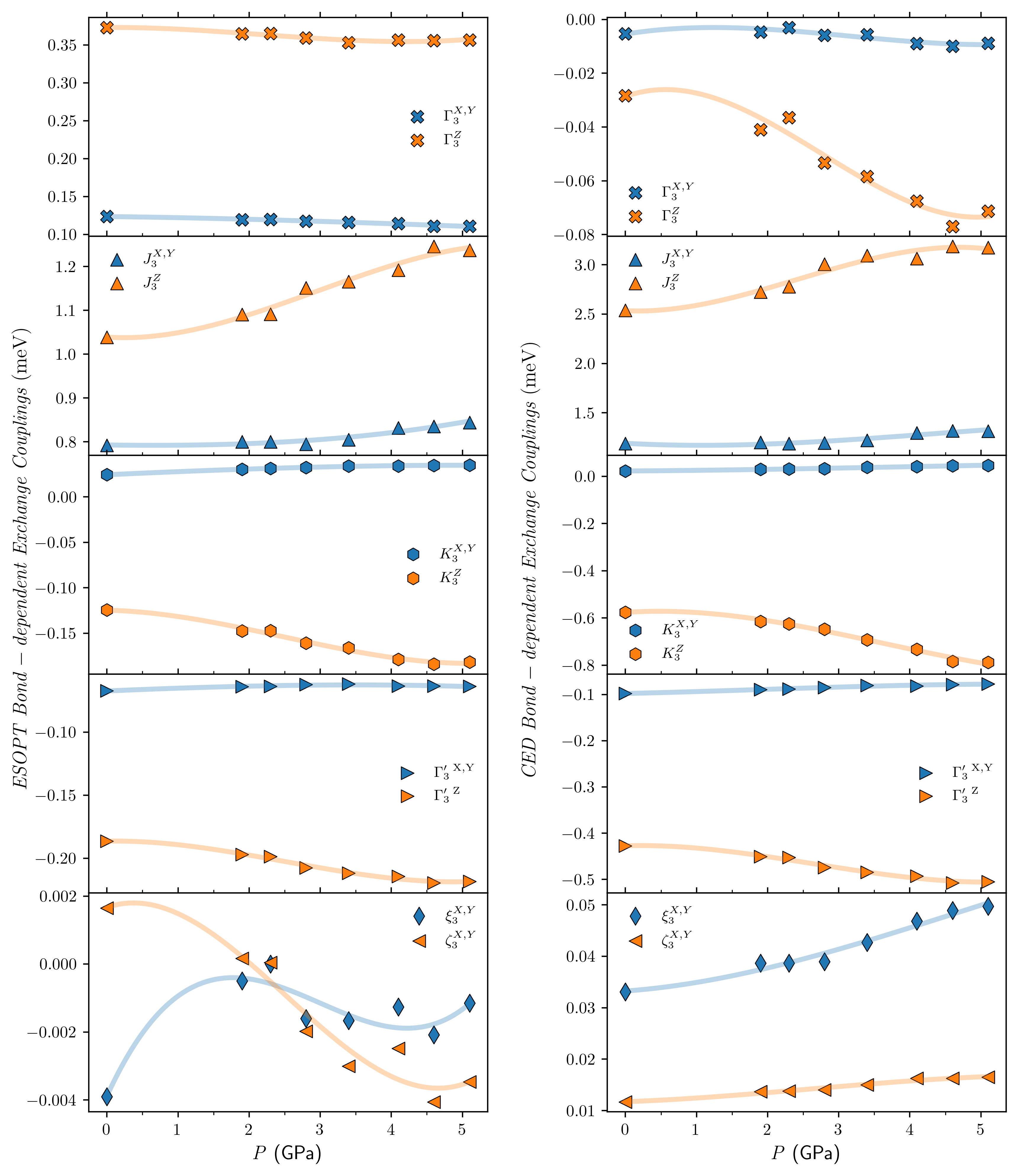}
  \caption{\label{fig:sm-3nn-esopt-ced}
  Pressure dependence of the complete set of third-neighbor exchange
  couplings obtained using exact second-order perturbation theory (ESOPT,
  left) and cluster exact diagonalization (CED, right). Blue and orange
  symbols denote the $X,Y$- and $Z$-bond values, respectively; solid curves
  are guides to the eye.}
\end{figure}

\section{Mean-field calculation of the ordering temperature}
\label{sec:sm-mean-field}
As a first approach to the magnetic ground state of the system, the mean-field technique is used.
Departing from equation (\ref{eq:couptens}), the mean-field decoupling can be written as:
\begin{equation}
\begin{aligned}
    \mathbf{S}_i^T \cdot \mathbb{J}_{ij} \cdot \mathbf{S}_j \approx{}& \mathbf{S}_i^T \cdot \mathbb{J}_{ij} \cdot \langle \mathbf{S}_j \rangle +
    \langle \mathbf{S}_i \rangle^T \cdot \mathbb{J}_{ij} \cdot \mathbf{S}_j- \langle \mathbf{S}_i \rangle^T \cdot \mathbb{J}_{ij} \cdot \langle \mathbf{S}_j \rangle
\end{aligned}
\end{equation}
taking the first term only, the mean field acting on spin $\mathbf{S}_i$ is $\mathbf{h}_i = \sum_j \mathbb{J}_{ij} \langle \mathbf{S}_j \rangle$ and can be written as:

\begin{equation}
     \mathbf{h}_i = \sum_n \mathbb{J}_{n}^\alpha \langle \mathbf{S}_j \rangle
\label{MeanF}     
\end{equation}
to generalize for $n$-th neighbors and $\alpha$-bonds.
Expressing eq. (\ref{MeanF}) in Fourier series for an ordering wavevector $\mathbf{Q}$, the total interaction tensor in momentum-space \citeS{SMKittel1996} is now:
\begin{equation}
    \mathbb{J}(\mathbf{Q}) = \sum_{n} \sum_{\delta_n} \mathbb{J}_{n}^\alpha e^{i\mathbf{Q} \cdot \mathbf{\delta}_n},
\end{equation}
where $\mathbf{\delta}^n_\alpha$ are the real space $n$-th neighbor $\alpha$-bond vectors.  The magnetic orderings to evaluate (FM, ZZ, etc)  enter the calculation here, as every magnetic ground state has a characteristic propagation vector $\mathbf{Q}$ \citeS{SMwills2001magnetic}. 

Next, the eigenvalues $\lambda$ of $\mathbb{J}(\mathbf{Q})$ are calculated and the smallest one is used for calculating the N\'eel temperature~\citeS{SMMaksi2019}:
\begin{equation}
    T_\text{N}^\text{MF}=-\frac{S(S+1)}{3k_B}\lambda_\text{min},
\end{equation}
where $\lambda_\text{min}$ is the lowest eigenvalue of the Fourier-transformed exchange matrix $\mathbb{J}_{ij}$ at the ordering vector $\mathbf{Q}$ \citeS{SMEhjal2004}.

Although our mean-field calculations do not reproduce the experimental values exactly \citeS{SMbahrami2022}, the reduction of $T_N$ with pressure is well captured qualitatively. Furthermore, the propagation vector $\mathbf{Q}=(0,0,0.5)$ corresponding to the A-type AFM order produces a higher $T_N$ at ambient pressure, surpassing that of the competing zigzag-ordered state ($\mathbf{Q}=(0,1/2,0)$).

\section{Classical analysis of FM and zigzag states}
\label{sec:sm-fm-zz-competition}

\subsection{Model}

We consider a layered honeycomb system with spins $S=1/2$ on the
vertices of each honeycomb layer. The model contains
$J$--$K$--$\Gamma$--$\Gamma'$ interactions, second- and third-neighbor
couplings $J_2$ and $J_3$, and an interlayer coupling $J_\perp$. We focus on
the two candidate states shown in Fig.~\ref{fig:sm-fm-vs-zz}: the FM state
and one of the three symmetry-related zigzag states.

\begin{figure}[t]
  \centering
  \includegraphics[width=0.6\textwidth]{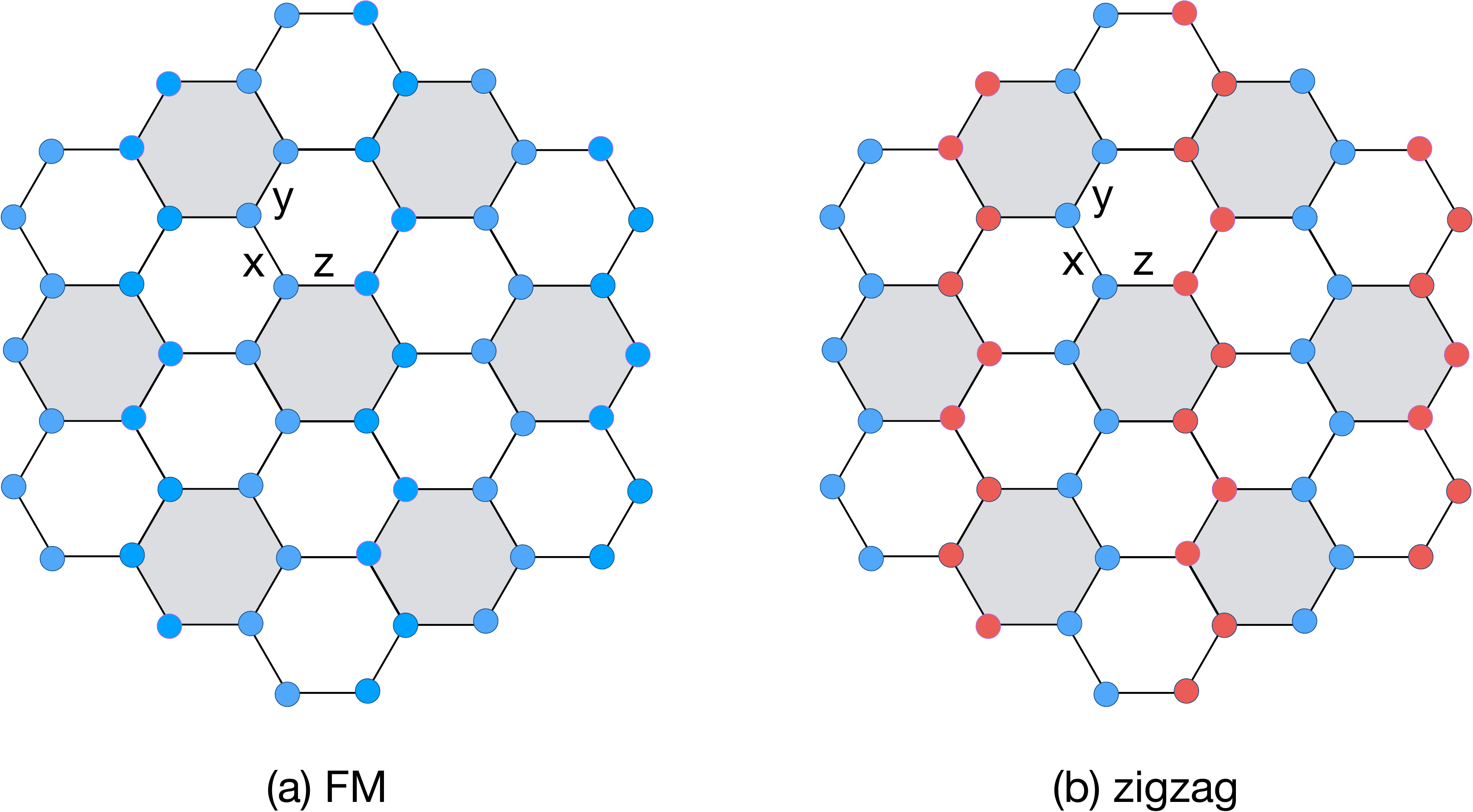}
  \caption{\label{fig:sm-fm-vs-zz}
  Ferromagnetic and zigzag states considered in the classical and mean-field
  analysis.}
\end{figure}

\subsection{FM state with alternating layers}

\subsubsection{Energy}

Consider a collinear state in which all spins within each layer point along
the same direction $\bm e=(e_x,e_y,e_z)$, while the direction alternates
between successive layers. In the classical limit, its energy is
\begin{equation}
E_{\mathrm{FM}}=E_{\mathrm{FM},K}+E_{\mathrm{FM},\Gamma}
+E_{\mathrm{FM},\Gamma'}+E_{\mathrm{FM},J}+E_{\mathrm{FM},J_\perp},
\end{equation}
where 
\begin{align}
E_{\mathrm{FM},K}
  &=\frac{N}{2}KS^2(e_x^2+e_y^2+e_z^2)=\frac{N}{2}KS^2,\\
E_{\mathrm{FM},\Gamma}
  &=N\Gamma S^2(e_xe_y+e_ye_z+e_ze_x),\\
E_{\mathrm{FM},\Gamma'}
  &=2N\Gamma'S^2(e_xe_y+e_ye_z+e_ze_x),\\
E_{\mathrm{FM},J}
  &=\frac{N}{2}S^2(3J_1+6J_2+3J_3-2J_\perp).
\end{align}
where N is the total number of sites.
Therefore,
\begin{equation}
\frac{E_{\mathrm{FM}}}{NS^2}
=\frac{1}{2}(K+3J_1+6J_2+3J_3-2J_\perp)
+(\Gamma+2\Gamma')(e_xe_y+e_ye_z+e_ze_x).
\label{eq:sm-fm-energy-general}
\end{equation}
The orientation-dependent factor can be written as
\begin{equation}
e_xe_y+e_ye_z+e_ze_x
=\frac{1}{2}(e_x+e_y+e_z)^2-\frac{1}{2},
\end{equation}
where $e_x^2+e_y^2+e_z^2=1$. Hence, when
$\Gamma+2\Gamma'<0$, the energy is minimized by maximizing
$(e_x+e_y+e_z)^2$, and the moments point along the cubic $[111]$ direction,
\begin{equation}
\bm e=\frac{1}{\sqrt{3}}(1,1,1).
\end{equation}
The minimum energy of the FM state with alternating layers is
\begin{equation}
\boxed{
\frac{E_{\mathrm{FM}}}{NS^2}
=\frac{1}{2}(K+3J_1+6J_2+3J_3-2J_\perp+2\Gamma+4\Gamma') }.
\label{eq:sm-fm-energy}
\end{equation}

For completeness, Eq.~\eqref{eq:sm-fm-energy-general} may also be written as
\begin{equation}
\frac{E_{\mathrm{FM}}}{NS^2}
=\frac{1}{2}(K+3J_1+6J_2+3J_3-2J_\perp)
+\frac{1}{2}\bm e^{\mathsf T}\bm\Lambda_{\mathrm{FM}}\bm e,
\end{equation}
with
\begin{equation}
\bm\Lambda_{\mathrm{FM}}=(\Gamma+2\Gamma')
\begin{pmatrix}
0&1&1\\
1&0&1\\
1&1&0
\end{pmatrix}.
\end{equation}
Its eigenvalues are
\begin{align}
\lambda_1&=2(\Gamma+2\Gamma'),\\
\lambda_2&=-(\Gamma+2\Gamma'),\\
\lambda_3&=\lambda_2,
\end{align}
with corresponding eigenvectors
\begin{align}
\bm v_1&=\frac{1}{\sqrt{3}}(1,1,1),\\
\bm v_2&=\frac{1}{\sqrt{2}}(1,0,-1),\\
\bm v_3&=\frac{1}{\sqrt{6}}(1,1,-2).
\end{align}
For $\Gamma+2\Gamma'<0$, $\lambda_1$ is the lowest eigenvalue, reproducing
Eq.~\eqref{eq:sm-fm-energy}.

\subsubsection{Ordering temperature}

The local mean field at site $i$ is
\begin{equation}
\bm h_i=h_i\bm e,\qquad h_i=\zeta\langle S_i\rangle,
\end{equation}
where
\begin{equation}
\boxed{\zeta=-(K+3J_1+6J_2+3J_3-2J_\perp+2\Gamma+4\Gamma')}.
\end{equation}
The self-consistency condition is
\begin{equation}
\langle S_i\rangle=S B_S\!\left(\frac{h_iS}{k_BT}\right),
\end{equation}
where
\begin{equation}
B_S(x)=\frac{2S+1}{2S}\coth\!\left(\frac{2S+1}{2S}x\right)
-\frac{1}{2S}\coth\!\left(\frac{x}{2S}\right).
\end{equation}
Equating the slopes at $\langle S_i\rangle=0$ gives
\begin{equation}
k_BT_N=\frac{S(S+1)}{3}\zeta.
\end{equation}
For $S=1/2$,
\begin{equation}
\boxed{
k_BT_N=\frac{1}{4}(-K-2\Gamma-4\Gamma'-3J_1-6J_2-3J_3+2J_\perp)}.
\label{eq:sm-fm-tn}
\end{equation}

\subsection{Zigzag state with alternating layers}

\subsubsection{Energy}

Consider the zigzag state with $X$ and $Y$ bonds FM and $Z$ bonds AFM
[Fig.~\ref{fig:sm-fm-vs-zz}(b)], again with alternating directions in
successive layers. Its classical energy is
\begin{equation}
E_{\mathrm{zz}}=E_{\mathrm{zz},K}+E_{\mathrm{zz},\Gamma}
+E_{\mathrm{zz},\Gamma'}+E_{\mathrm{zz},J}+E_{\mathrm{zz},J_\perp},
\end{equation}
with
\begin{align}
E_{\mathrm{zz},K}
  &=\frac{N}{2}KS^2(e_x^2+e_y^2-e_z^2),\\
E_{\mathrm{zz},\Gamma}
  &=N\Gamma S^2(e_ye_z+e_ze_x-e_xe_y),\\
E_{\mathrm{zz},\Gamma'}
  &=2N\Gamma'S^2e_xe_y,\\
E_{\mathrm{zz},J}
  &=\frac{N}{2}S^2(J_1-2J_2-3J_3-2J_\perp).
\end{align}
Thus,
\begin{align}
\frac{E_{\mathrm{zz}}}{NS^2}
=\frac{1}{2}\big[&K(e_x^2+e_y^2-e_z^2)
+2\Gamma(e_ye_z+e_ze_x-e_xe_y)\nonumber\\
&+4\Gamma'e_xe_y+J_1-2J_2-3J_3-2J_\perp\big].
\end{align}
Equivalently,
\begin{equation}
\frac{E_{\mathrm{zz}}}{NS^2}
=\frac{1}{2}\bm e^{\mathsf T}\bm\Lambda_{\mathrm{zz}}\bm e
+\frac{1}{2}(J_1-2J_2-3J_3-2J_\perp),
\end{equation}
where
\begin{equation}
\bm\Lambda_{\mathrm{zz}}=
\begin{pmatrix}
K&-\Gamma+2\Gamma'&\Gamma\\
-\Gamma+2\Gamma'&K&\Gamma\\
\Gamma&\Gamma&-K
\end{pmatrix}.
\end{equation}
The eigenvalues are
\begin{align}
\lambda_1&=\Gamma-2\Gamma'+K,\\
\lambda_2&=-\frac{1}{2}\Gamma+\Gamma'-\frac{1}{2}c,\\
\lambda_3&=-\frac{1}{2}\Gamma+\Gamma'+\frac{1}{2}c,
\end{align}
where
\begin{equation}
c=\sqrt{8\Gamma^2+[\Gamma-2(\Gamma'+K)]^2}.
\end{equation}
In the parameter range of interest, $\lambda_2$ is the lowest eigenvalue.
The corresponding eigenvector is
\begin{equation}
\bm v_2=\frac{1}{\sqrt{2a^2+b^2}}(a,a,b),
\end{equation}
with
\begin{align}
a={}&-3\Gamma^2+2\Gamma'(c-2\Gamma'-2K)
+\Gamma(-c+4\Gamma'+2K),\\
b={}&\Gamma(c+3\Gamma-6\Gamma'+2K),
\end{align}
which satisfy
\begin{equation}
\boxed{1+2\frac{a}{b}-\frac{b}{a}=2\frac{\Gamma'+K}{\Gamma}}.
\end{equation}
The minimum energy is therefore
\begin{equation}
\boxed{\begin{aligned}
\frac{E_{\mathrm{zz}}}{NS^2}
=\frac{1}{2}\bigg[&-\frac{1}{2}\Gamma+\Gamma'
-\frac{1}{2}\sqrt{8\Gamma^2+[\Gamma-2(\Gamma'+K)]^2}\\
&+J_1-2J_2-3J_3-2J_\perp\bigg]
\end{aligned}}
\label{eq:sm-zz-energy}
\end{equation}

\subsubsection{Ordering temperature}

Let the blue and red sites in Fig.~\ref{fig:sm-fm-vs-zz} correspond to
$+\bm e$ and $-\bm e$, respectively. The local field on a blue site is
\begin{align}
\bm h_1={}&-K(e_x,e_y,-e_z)
-\Gamma(e_z-e_y,e_z-e_x,e_x+e_y)\nonumber\\
&-2\Gamma'(e_y,e_x,0)
+(-J_1+2J_2+3J_3+2J_\perp)\bm e.
\end{align}
Its components obey
\begin{align}
\frac{h_{1x}}{e_x}=\frac{h_{1y}}{e_y}=\frac{h_{1z}}{e_z}
=\frac{1}{2}\big[&\Gamma-2\Gamma'-2J_1+4J_2+6J_3+4J_\perp\nonumber\\
&+\sqrt{8\Gamma^2+[\Gamma-2(\Gamma'+K)]^2}\big].
\end{align}
Hence $\bm h_1=\zeta_{\mathrm{zz}}\langle S_1\rangle\bm e$, with
\begin{align}
\zeta_{\mathrm{zz}}=\frac{1}{2}\big[&\Gamma-2\Gamma'-2J_1
+4J_2+6J_3+4J_\perp\nonumber\\
&+\sqrt{8\Gamma^2+[\Gamma-2(\Gamma'+K)]^2}\big].
\end{align}
The zigzag ordering temperature is
\begin{equation}
\boxed{\begin{aligned}
k_BT_{\mathrm{zz},N}=\frac{S(S+1)}{6}\big[&\Gamma-2\Gamma'-2J_1
+4J_2+6J_3+4J_\perp\\
&+\sqrt{8\Gamma^2+[\Gamma-2(\Gamma'+K)]^2}\big]
\end{aligned}}
\label{eq:sm-zz-tn}
\end{equation}

\subsection{Comparison between FM and zigzag states}

The transition between the two phases occurs when
$\Delta E=E_{\mathrm{FM}}-E_{\mathrm{zz}}=0$, or equivalently when
\begin{equation}
\boxed{
K+\frac{5}{2}\Gamma+3\Gamma'
+\frac{1}{2}\sqrt{8\Gamma^2+[\Gamma-2(\Gamma'+K)]^2}
+2J_1+8J_2+6J_3=0 }
\label{eq:sm-fm-zz-boundary}
\end{equation}
Figure~\ref{fig:sm-energy-fm-vs-zz} shows this transition line for
$J_1=J_2=0$ and $J_3=0.2|K|$.

\begin{figure}[t]
  \centering
  \includegraphics[width=0.6\textwidth]{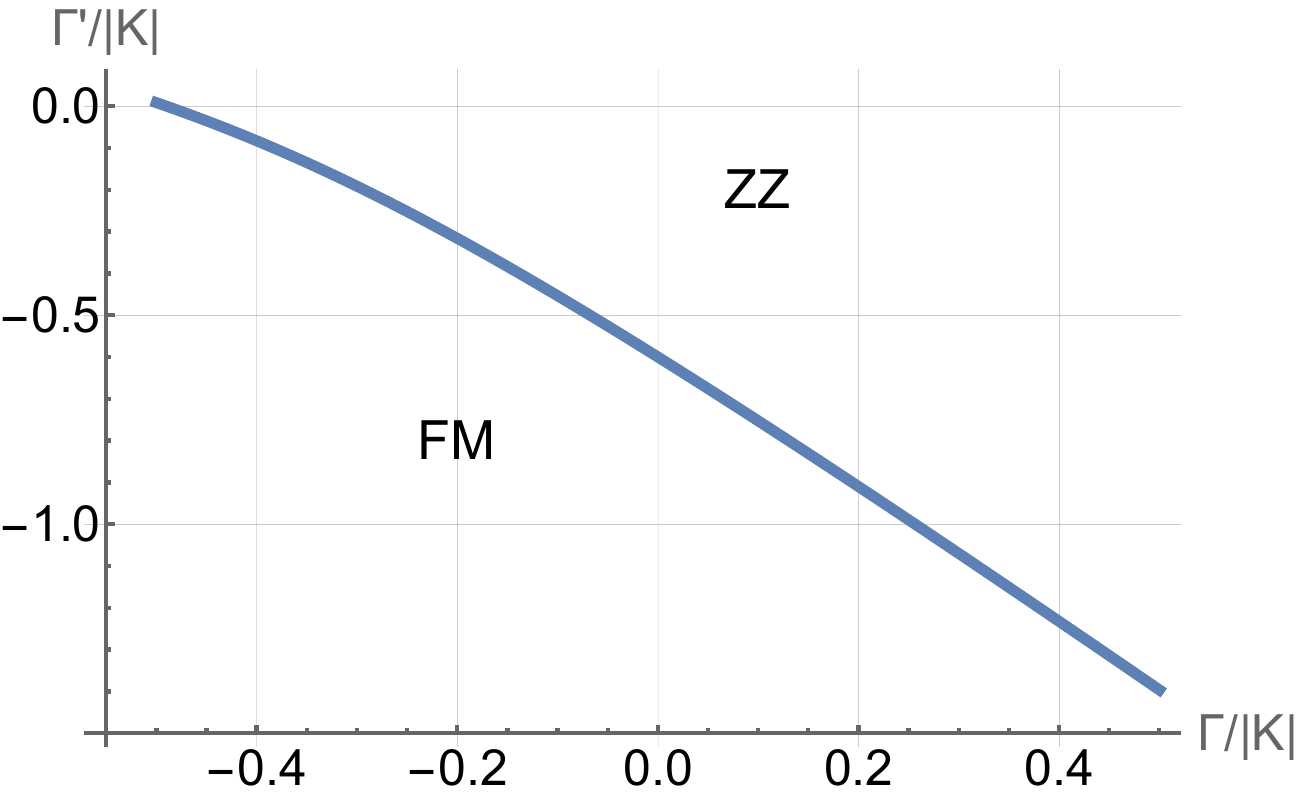}
  \caption{\label{fig:sm-energy-fm-vs-zz}
  Transition line between the FM and zigzag states for $J_1=J_2=0$ and
  $J_3=0.2|K|$.}
\end{figure}

Figure~\ref{fig:sm-energy-fm-vs-zz-j3} shows the transition lines for
several values of $J_3/|K|$. The lines shift downward as $J_3$ increases,
consistent with third-neighbor exchange favoring zigzag order.

\begin{figure}[t]
  \centering
  \includegraphics[width=0.6\textwidth]{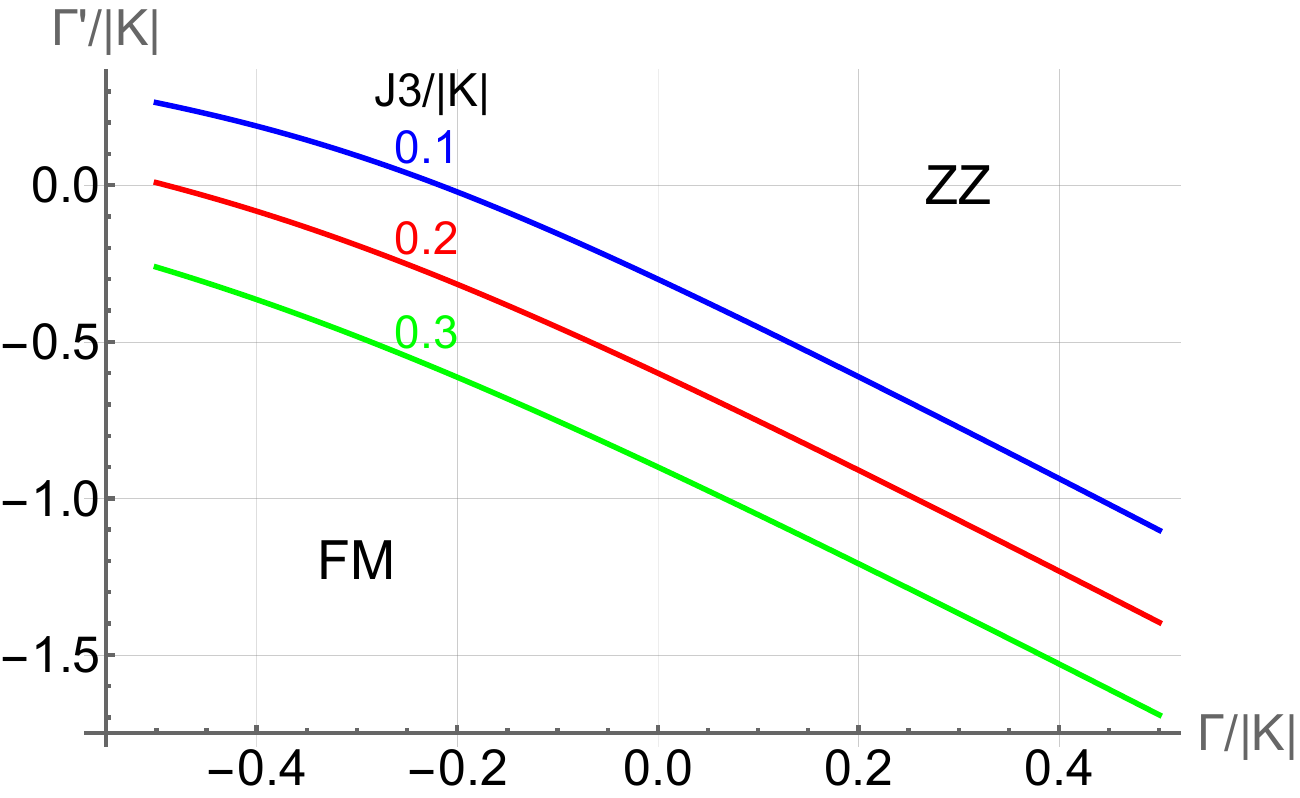}
  \caption{\label{fig:sm-energy-fm-vs-zz-j3}
  Transition lines between the FM and zigzag states for several values of
  $J_3/|K|$, with $J_1=J_2=0$.}
\end{figure}

Figure~\ref{fig:sm-tn-contour} shows a contour plot of the FM ordering
temperature for $J_3=0.2|K|$, $K=-10$~meV, and $J_1=J_2=0$.

\begin{figure}[t]
  \centering
  \includegraphics[width=0.6\textwidth]{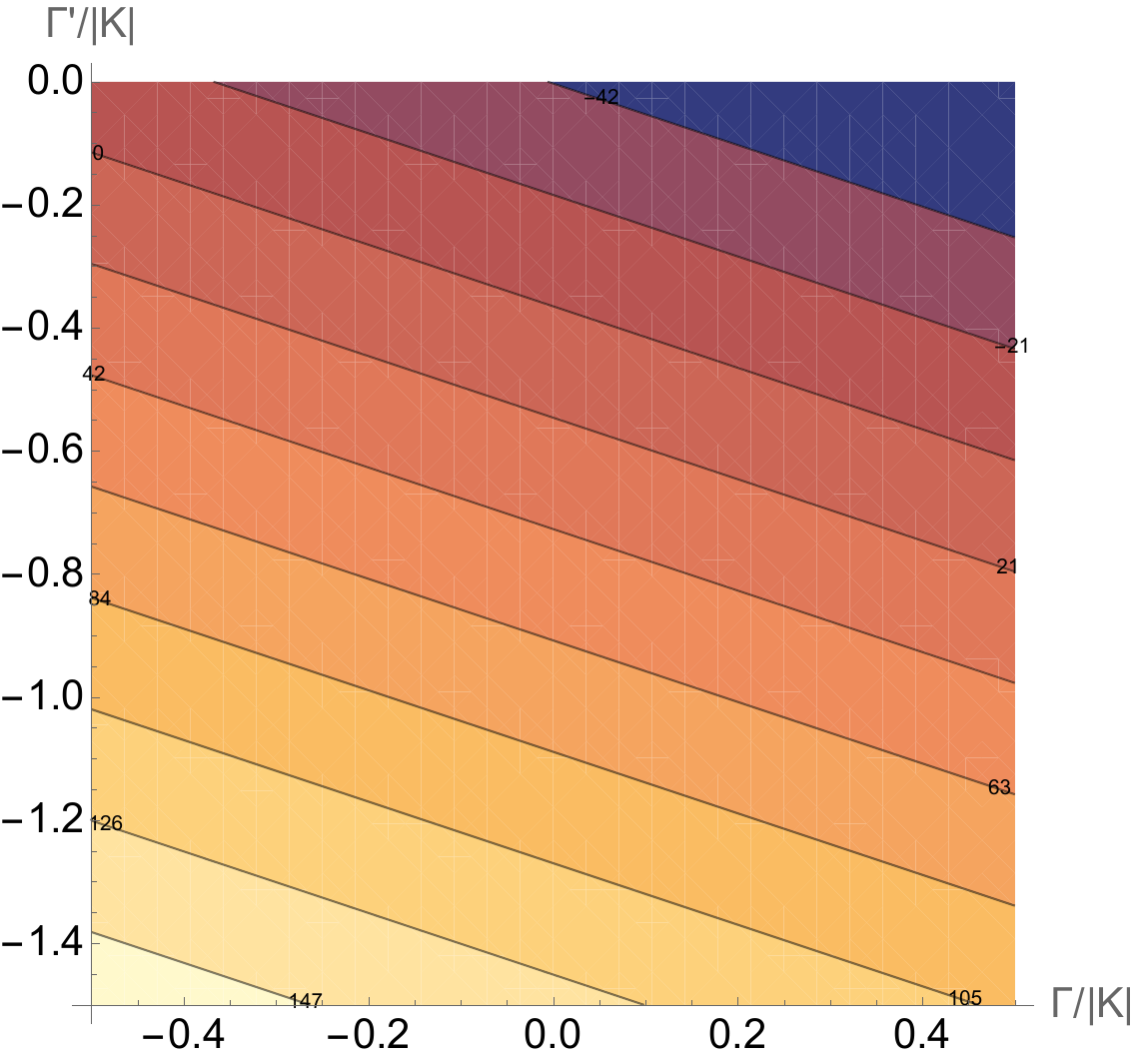}
  \caption{\label{fig:sm-tn-contour}
  Mean-field ordering temperature of the FM phase for $J_3=0.2|K|$,
  $K=-10$~meV, and $J_1=J_2=0$.}
\end{figure}

\newpage
\section{DMRG results}
\label{sec:sm-dmrg}

\begin{figure}[b]
\centering
\includegraphics[width=\textwidth]{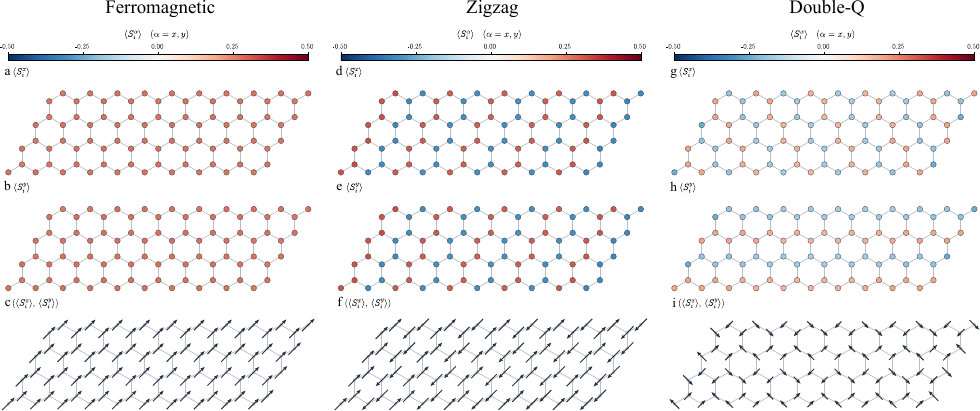}
\caption{The bulk magnetic textures obtained by DMRG. The columns show FM [(a)--(c)], zigzag [(d)--(f)], and double-$Q$ [(g)--(i)] states. The first and second rows show $m_i^x=\langle S_i^x\rangle$ and $m_i^y=\langle S_i^y\rangle$: red and blue denote positive and negative
components, respectively. The third row shows the directions of the local magnetic moments $(m_i^x,m_i^y)$. Only the bulk is displayed, with both open ends omitted.}
\label{fig:sm-dmrg-moments}
\end{figure}

We investigate the spin-$1/2$ model using the density-matrix
renormalization group (DMRG)~\citeS{SMWhite1992,SMSchollwock2011,SMtenpy2024} on a honeycomb YC4 cylinder without particle-number conservation, with its circumference containing 8 sites and a cylinder length $L_x=12$. Boundary conditions are periodic around the circumference and open along the cylinder axis. 
The simulation used the maximum bond dimensions $900$ and $1600$, with the largest truncation error of order $10^{-8}$ and $10^{-9}$, respectively. Both obtain consistent results for the magnetic patterns and correlations. 

We scan the two nearest-neighbor symmetric off-diagonal exchanges
$\Gamma_1$ and $\Gamma_1'$ while holding the other Hamiltonian parameters, $K_1$ and $J_3$, fixed to $-1$ and $0.2$ respectively. Let $E_0(\Gamma_1,\Gamma_1')$ denote the optimized ground-state energy and $e_0=E_0/N$ the energy per spin, where $N$ is the number of sites in the calculation. The exchange susceptibilities are defined by
\begin{equation}
\chi_{\Gamma_1}=-\left.
\frac{\partial^2 e_0}{\partial\Gamma_1^2}\right|_{\Gamma_1'},
\qquad
\chi_{\Gamma_1'}=-\left.
\frac{\partial^2 e_0}{\partial(\Gamma_1')^2}\right|_{\Gamma_1}.
\label{eq:sm-dmrg-susceptibility}
\end{equation}
which measure the response to an exchange coupling. We evaluate these by numerical differentiation of the energies on the coupling grid, together with the correlations and static structure factors, to identify the phase boundaries. 
Figure~\ref{fig:sm-dmrg-moments} compares representative ferromagnetic $(K_1=-1,\Gamma_1=-0.3,\Gamma_1'=-0.4,J_3=0.2)$, zigzag $(K_1=-1,\Gamma_1=-0.1,\Gamma_1'=-0.3,J_3=0.2)$, and double-$Q$ $(K_1=-1,\Gamma_1=-0.1,\Gamma_1'=0.0,J_3=0.2)$ states by magnetizations.

\begin{figure}[htbp]
\centering
\includegraphics[width=\textwidth]{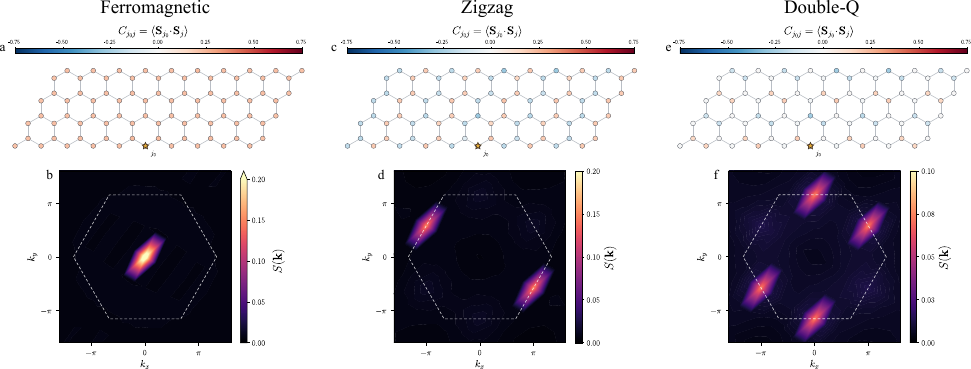}
\caption{Spin correlations and static spin structure factors for
the same three states as in Fig.~\ref{fig:sm-dmrg-moments}.
(a), (c), and (e): full dot-product correlation
$C_{j_0j}=\langle\bm S_{j_0}\cdot\bm S_j\rangle$ for the FM, zigzag, and double-$Q$ states, respectively. Only the bulk sites are shown; The star marks the bulk reference site
$j_0$. (b), (d), and (f): magnitudes of the structure factor using reference-site Fourier transforms,
$S(\bm k)=|\mathcal F_{j_0}(\bm k)|$, with the $1/N_{\mathrm c}$
normalization of Eq.~\eqref{eq:fourier}. Dashed hexagons mark the first Brillouin zone. The dominant weight lies at the zone center for FM order, at one inequivalent $M$ point for zigzag order, and at two inequivalent $M$ points for the double-$Q$ texture. }
\label{fig:sm-dmrg-correlations}
\end{figure}

We also computed the equal-time spin correlation, defined as
\begin{equation}
C_{ij}=\langle\bm S_i\cdot\bm S_j\rangle
=\sum_{\alpha=x,y,z}\langle S_i^\alpha S_j^\alpha\rangle.
\label{eq:sm-dmrg-correlation}
\end{equation}
The upper row of Fig.~\ref{fig:sm-dmrg-correlations} fixes the first site to the starred bulk reference site $j_0$ and displays $C_{j_0j}$ on the bulk sites $j$, excluding both open ends. Red and blue indicate positive and negative correlations with $j_0$. 
The static structure factors are obtained from the same reference-site correlations, with normalization $1/N_{\mathrm c}$:
\begin{equation}
\mathcal F_{j_0}(\bm k)=\frac{1}{N_{\mathrm c}}
\sum_{j\in\mathcal W}e^{i\bm k\cdot(\bm r_j-\bm r_{j_0})}C_{j_0j}.
\label{eq:fourier}
\end{equation}
Here $\mathcal W$ is the set of bulk sites included in the Fourier sum, $N_{\mathrm c}=|\mathcal W|$, and $\bm r_j$ is the position of site $j$ in the cylinder honeycomb lattice.

\section{Linear spin-wave theory}
\label{sec:sm-spin-waves}
We consider a periodic honeycomb layer of $N=2N_c$ spin-$S$ moments, with $N_c$ two-site unit cells and $\hbar=1$. The relevant generic exchange Hamiltonian is
\begin{equation}\label{eq:model}
H=\sum_{\langle ij\rangle_\gamma}\bm S_i^{\mathsf T}\mathbb J_\gamma\bm S_j
 +J_3\sum_{\langle ij\rangle_3}\bm S_i\cdot\bm S_j,
\;\; \gamma=X,Y,Z.
\end{equation}
The sums run over nearest neighbors and third neighbors across a hexagon, respectively. In cubic spin axes,
the nearest-neighbor exchange tensors are
\begin{equation}\label{eq:em-tensors}
\mathbb J_X=\begin{pmatrix}J+K&\Gamma'&\Gamma'\\\Gamma'&J&\Gamma\\\Gamma'&\Gamma&J\end{pmatrix},\;\;
\mathbb J_Y=\begin{pmatrix}J&\Gamma'&\Gamma\\\Gamma'&J+K&\Gamma'\\\Gamma&\Gamma'&J\end{pmatrix},\;\;
\mathbb J_Z=\begin{pmatrix}J&\Gamma&\Gamma'\\\Gamma&J&\Gamma'\\\Gamma'&\Gamma'&J+K\end{pmatrix}.
\end{equation}
Here $J$ and $J_3$ are isotropic exchanges, $K$ is the Kitaev exchange, and $\Gamma,\Gamma'$ are symmetric off-diagonal exchanges (for simplicity, here we suppress the nearest-neighbor subscript $\circ_1$ for $K_1$, $\Gamma_1$ and $\Gamma_1'$ used in the main text). 

We expand about the uniform ferromagnetic configuration
$\bm S_i=S\bm n$, with $\bm n=(1,1,1)/\sqrt3$.
Its classical energy $E_{\rm cl}$ is the expectation value of $H$ in the product state with each spin fully polarized along $\bm n$, which, on a $Z$ bond, for example, gives
\begin{equation}\label{eq:em-classical-bond}
\mathcal E_Z^{\rm cl}=S^2\bm n^{\mathsf T}\mathbb J_Z\bm n
 =\frac{S^2}{3}\big[J+J+(J+K)+2\Gamma+4\Gamma'\big]
 =S^2\left[J+\frac{K+2\Gamma+4\Gamma'}3\right].
\end{equation}
Cyclic permutation gives the same energy for $X$ and $Y$ bonds. There are $N/2=N_c$ bonds of each nearest-neighbor type and $3N/2=3N_c$ third-neighbor bonds: each site has one neighbor of each type and three third neighbors, and each bond has two endpoints. Consequently,
\begin{equation}\label{eq:em-classical}
E_{\rm cl}=N_c\big(\mathcal E_X^{\rm cl}+\mathcal E_Y^{\rm cl}
+\mathcal E_Z^{\rm cl}\big)+3N_cJ_3S^2
=N_cS^2(3J+K+2\Gamma+4\Gamma'+3J_3).
\end{equation}

To describe fluctuations around this configuration, we use a common
local frame whose third axis lies along the ordered moment. Choose $\bm e_3=\bm n$ perpendicular to the honeycomb plane, $\bm e_1=(1,-1,0)/\sqrt2$ perpendicular to $\bm e_3$,
and $\bm e_2=\bm e_3\times\bm e_1=(1,1,-2)/\sqrt6$.
These give an orthonormal matrix $R$ relating the two frames:
\begin{equation}\label{eq:em-frame}
R=(\bm e_1,\bm e_2,\bm e_3)=
\begin{pmatrix}
1/\sqrt2&1/\sqrt6&1/\sqrt3\\
-1/\sqrt2&1/\sqrt6&1/\sqrt3\\
0&-2/\sqrt6&1/\sqrt3
\end{pmatrix},\;\;
\bm S_i=R\widetilde{\bm S}_i,\;\;
\widetilde{\bm S}_i=R^{\mathsf T}\bm S_i.
\end{equation}
The local components are given by its projections
$\widetilde S_i^\mu=\bm e_\mu\cdot\bm S_i$; in particular, the reference configuration along
$\widetilde S_i^3=(S_i^x+S_i^y+S_i^z)/\sqrt3$ has local components $(0,0,S)$. Substitution of $\bm S_i=R\widetilde{\bm S}_i$ on both sites of a bond rotates its exchange tensor as
\begin{equation}\label{eq:em-rotated-tensor}
\bm S_i^{\mathsf T}\mathbb J_\gamma\bm S_j
=\widetilde{\bm S}_i^{\mathsf T} \mathbb L_\gamma\widetilde{\bm S}_j,
\;\; \mathbb L_\gamma=R^{\mathsf T}\mathbb J_\gamma R,
\;\; \mathbb L_{\gamma,\mu\nu}=\bm e_\mu^{\mathsf T}\mathbb J_\gamma\bm e_\nu.
\end{equation}
Thus the indices $\mu,\nu=1,2,3$ specify local spin components, and
$\mathbb L_{\gamma,33}$ is the longitudinal exchange coefficient on a
$\gamma$ bond. Direct matrix multiplication yields
\begin{equation}\label{eq:em-rotated-z}
\mathbb L_Z=\begin{pmatrix}
J-\Gamma&0&0\\
0&J+(2K+\Gamma-4\Gamma')/3&\sqrt2(\Gamma-K-\Gamma')/3\\
0&\sqrt2(\Gamma-K-\Gamma')/3&J+(K+2\Gamma+4\Gamma')/3
\end{pmatrix}.
\end{equation}
Applying the same rotation to the other two bond types gives
\begin{equation}\label{eq:em-rotated-x}
\mathbb L_X=\begin{pmatrix}
J+K/2-\Gamma' & \dfrac{K+2\Gamma-2\Gamma'}{2\sqrt3} & \dfrac{K-\Gamma+\Gamma'}{\sqrt6}\\[5pt]
\dfrac{K+2\Gamma-2\Gamma'}{2\sqrt3} & J+K/6-2\Gamma/3-\Gamma'/3 & \dfrac{K-\Gamma+\Gamma'}{3\sqrt2}\\[5pt]
\dfrac{K-\Gamma+\Gamma'}{\sqrt6} & \dfrac{K-\Gamma+\Gamma'}{3\sqrt2} & J+(K+2\Gamma+4\Gamma')/3
\end{pmatrix},
\end{equation}
\begin{equation}\label{eq:em-rotated-y}
\mathbb L_Y=\begin{pmatrix}
J+K/2-\Gamma' & -\dfrac{K+2\Gamma-2\Gamma'}{2\sqrt3} & -\dfrac{K-\Gamma+\Gamma'}{\sqrt6}\\[5pt]
-\dfrac{K+2\Gamma-2\Gamma'}{2\sqrt3} & J+K/6-2\Gamma/3-\Gamma'/3 & \dfrac{K-\Gamma+\Gamma'}{3\sqrt2}\\[5pt]
-\dfrac{K-\Gamma+\Gamma'}{\sqrt6} & \dfrac{K-\Gamma+\Gamma'}{3\sqrt2} & J+(K+2\Gamma+4\Gamma')/3
\end{pmatrix}.
\end{equation}
where all three tensors have the same $33$ entry, while their
transverse entries retain the bond dependence that generates boson pairing.
Isotropic third-neighbor exchange remains $R^{\mathsf T}(J_3\mathbb I_3)R=J_3\mathbb I_3$.

Introducing $[a_i,a_j^\dagger]=\delta_{ij}$ and
$\widehat n_i=a_i^\dagger a_i$, the Holstein--Primakoff expansion about
this local third axis is
\begin{equation}\label{eq:hp}
\widetilde S_i^3=S-\widehat n_i,\;\;
\widetilde S_i^1\simeq\sqrt{\frac S2}(a_i+a_i^\dagger),\;\;
\widetilde S_i^2\simeq-i\sqrt{\frac S2}(a_i-a_i^\dagger).
\end{equation}
Linear spin-wave theory retains the constant and quadratic terms,
$H_{\rm LSWT}=E_{\rm cl}+H_2$.

The mixed longitudinal--transverse entries of $\mathbb L_\gamma$ generate linear and cubic terms. The linear terms cancel in the lattice sum; the cubic terms are beyond the
harmonic approximation.
For the longitudinal product,
$\mathbb L_{\gamma,33}\widetilde S_i^3\widetilde S_j^3
=\mathbb L_{\gamma,33}[S^2-S(\widehat n_i+\widehat n_j)+\widehat n_i\widehat n_j]$.
Discarding its quartic term and collecting the transverse products gives
\begin{equation}\label{eq:rotation}
\begin{aligned}
H_{ij}^{(2)}&=-S\mathbb L_{\gamma,33}(\widehat n_i+\widehat n_j)
 +\big[t_\gamma a_i^\dagger a_j+p_\gamma a_i^\dagger a_j^\dagger+\mathrm{H.c.}\big],\\
t_\gamma&=\frac S2(\mathbb L_{\gamma,11}+\mathbb L_{\gamma,22}),\;\;
p_\gamma=\frac S2(\mathbb L_{\gamma,11}-\mathbb L_{\gamma,22}+2i\mathbb L_{\gamma,12}).
\end{aligned}
\end{equation}
with $t_\gamma$ describing boson hopping and $p_\gamma$ boson pair creation. The phases of $p_\gamma$ are fixed by the transverse axes $\bm e_1,\bm e_2$. 
Evaluating the rotated tensors gives $t_X=t_Y=t_Z=t_1$ and
\begin{equation}\label{eq:coeff}
\begin{aligned}
\epsilon_0&=-S(3J+K+2\Gamma+4\Gamma'+3J_3),&
Q&=K+2\Gamma-2\Gamma',\\
t_1&=S\left[J+\frac{K-\Gamma-2\Gamma'}3\right],&
t_3&=SJ_3,\\
p_Z&=-\frac{SQ}{3},&
p_X&=\frac{SQ}{6}(1+i\sqrt3),\;\; p_Y=p_X^*.
\end{aligned}
\end{equation}
The quadratic real-space Hamiltonian is therefore
\begin{equation}\label{eq:realspace}
H_2=\epsilon_0\sum_i\widehat n_i
 +\sum_{\langle ij\rangle_\gamma}
 \big[t_1a_i^\dagger a_j+p_\gamma a_i^\dagger a_j^\dagger+\mathrm{H.c.}\big]
 +t_3\sum_{\langle ij\rangle_3}(a_i^\dagger a_j+\mathrm{H.c.}).
\end{equation}
Now we express the quadratic Hamiltonian in momentum space. Let the nearest-neighbor distance be $1$, so that the vectors from an $A$ site to its three $B$ neighbors and the primitive vectors for the Bravais lattice are
\begin{equation}\label{eq:em-lattice}
\begin{gathered}
\bm\delta_Z=(0,-1),\;\;
\bm\delta_X=(-\sqrt3/2,1/2),\;\;
\bm\delta_Y=(\sqrt3/2,1/2),\\
\bm a_1=\bm\delta_Z-\bm\delta_X=(\sqrt3/2,-3/2),\;\;
\bm a_2=\bm\delta_Y-\bm\delta_Z=(\sqrt3/2,3/2).
\end{gathered}
\end{equation}
A cell at $\bm R=m\bm a_1+n\bm a_2$ contains $A$ at $\bm R$ and $B$ at $\bm R+\bm\delta_Z$. The $Z,Y,X$ nearest-neighbor bonds connect
$A_{\bm R}$ to $B_{\bm R}$, $B_{\bm R+\bm a_2}$, and
$B_{\bm R-\bm a_1}$, respectively. The three third-neighbor cell shifts in the Bravais lattice are $\bm a_1+\bm a_2$, $-\bm a_1-\bm a_2$, and
$-\bm a_1+\bm a_2$, corresponding to physical displacements
$-2\bm\delta_X$, $-2\bm\delta_Y$, and $-2\bm\delta_Z$.
For sublattice $s=A,B$, define
\begin{equation}\label{eq:em-fourier}
a_{\bm R s}=\frac1{\sqrt{N_c}}\sum_{\bm q}e^{i\bm k\cdot\bm R}a_{s\bm q},
\;\; q_\mu=\bm k\cdot\bm a_\mu.
\end{equation}
The two dimensionless coordinates $q_1,q_2$ span a full reciprocal
primitive cell. Fourier transformation of Eq.~\eqref{eq:realspace}
then yields the hopping and pair functions
\begin{equation}\label{eq:fg}
\begin{aligned}
f(\bm q)&=t_1(1+e^{iq_2}+e^{-iq_1})
 +t_3\big[2\cos(q_1+q_2)+e^{i(q_2-q_1)}\big],\\
g_+(\bm q)&=p_Z+p_Ye^{iq_2}+p_Xe^{-iq_1},\;\;
g_-(\bm q)=p_Z+p_Ye^{-iq_2}+p_Xe^{iq_1}.
\end{aligned}
\end{equation}
Here $g_-(\bm q)=g_+(-\bm q)$. The phases follow from the cell shifts
specified above, with $f(-\bm q)=f(\bm q)^*$ for real hopping amplitudes.
Collecting the two sublattice annihilation operators and their
creation partners defines the Nambu spinor
\begin{equation}\label{eq:em-nambu}
\Psi_{\bm q}=
(a_{A\bm q},a_{B\bm q},a^\dagger_{A,-\bm q},a^\dagger_{B,-\bm q})^{\mathsf T},
\;\; \Sigma_3=\operatorname{diag}(1,1,-1,-1).
\end{equation}
Positive definiteness of $\mathcal H$ throughout the Brillouin zone establishes local harmonic (meta)stability. 
With the coefficients in Eq.~\eqref{eq:coeff} and functions in Eq.~\eqref{eq:fg}, the compact form of LSWT is:
\begin{equation}\label{eq:bdg}
H_{\rm LSWT}=E_{\rm cl}-N_c\epsilon_0
 +\frac12\sum_{\bm q}\Psi_{\bm q}^\dagger\mathcal H(\bm q)\Psi_{\bm q},
\end{equation}
\begin{equation}
\mathcal H(\bm q)=
\begin{pmatrix}
\epsilon_0&f(\bm q)&0&g_+(\bm q)\\
f(\bm q)^*&\epsilon_0&g_-(\bm q)&0\\
0&g_-(\bm q)^*&\epsilon_0&f(\bm q)\\
g_+(\bm q)^*&0&f(\bm q)^*&\epsilon_0
\end{pmatrix}
\end{equation}
Diagonalizing with a para-unitary transformation and using $S = 1/2$ give the linear spin wave results shown in the main text. Chern numbers are obtained from a discretized Berry curvature following \citeS{SMFukui05}.

\begin{figure}[!t]
  \includegraphics{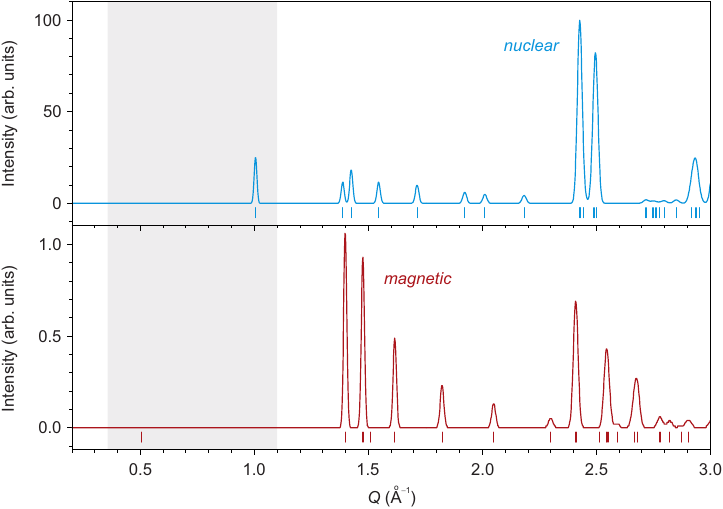}
  \caption{\label{fig:sm-neutron-patterns}
  Simulated nuclear and magnetic scattering for the A-AFM order in Ag$_3$LiRh$_2$O$_6$ with the spins perpendicular to the honeycomb planes and $\mu_{\rm Rh}=1$\,$\mu_B$. The shaded region shows the $Q$-range probed in Ref.~\protect\citeS{SMSakrikar2025}.}
\end{figure}

\section{Simulation of the neutron diffraction pattern}

Choosing the ground state anticipated from our microscopic magnetic model, we simulated neutron diffraction pattern of Ag$_3$LiRh$_2$O$_6$ using Mag2Pol \citeS{SMQureshi:in5014}. The A-type AFM order corresponds to the propagation vector $\mathbf k=(0,0,\frac12)$ and magnetic space group $C_c\,2/c$ for spins perpendicular to the honeycomb planes and, therefore, parallel to the $\mathbf c^*$ direction of the reciprocal lattice. Fig.~\ref{fig:sm-neutron-patterns} displays nuclear scattering, as well as magnetic scattering calculated assuming the ordered moment of 1\,$\mu_B$ on the pseudospin-$\frac12$ Rh$^{4+}$ ion. 

Magnetic Bragg peaks of Ag$_3$LiRh$_2$O$_6$ are expected to reach about 1\% of the maximum nuclear intensity. Two of them, $(0\,2\,\frac12)$ at $Q\simeq 1.48$\,\r A$^{-1}$ and $(1\,1\,-\frac12)$ at $Q\simeq 1.62$\,\r A$^{-1}$, do not overlap with the nuclear peaks and should be observable on a high-flux instrument. An interesting feature of the proposed A-AFM state of the $K_1-\Gamma_1'$ model is the complete suppression of the low-angle reflection $(0\,0\,\frac12)$ at $Q\simeq 0.50$\,\r A$^{-1}$ because its $Q$-vector is parallel to the magnetic moment. The magnetic Bragg peaks of Ag$_3$LiRh$_2$O$_6$ thus lie at higher angles than the first nuclear Bragg peak, at odds with the typical magnetic scattering. The neutron experiment on Ag$_3$LiRh$_2$O$_6$ reported in Ref.~\citeS{SMSakrikar2025} was restricted to the $0.35-1.1$\,\r A$^{-1}$ range and did not cover the $Q$-range where observable magnetic Bragg peaks are expected, hence no magnetic scattering could be detected. A neutron diffraction experiment focusing on the $Q=1-2$\,\r A$^{-1}$ range would be an interesting direction for future studies and a direct experimental proof of the proposed A-AFM order in Ag$_3$LiRh$_2$O$_6$.


\end{widetext}

\bibliographystyleS{apsrev4-2}

\renewcommand{\bibnumfmt}[1]{[S#1]}

\end{document}